\documentclass[10pt,journal,compsoc]{IEEEtran}
\usepackage{svg}
\usepackage{py}
\usepackage[skip=0pt]{caption}

\usepackage{fancyhdr}
\usepackage[normalem]{ulem}
\usepackage[hyphens]{url}
\usepackage[sort,nocompress]{cite}
\usepackage[final]{microtype}
\usepackage[keeplastbox]{flushend}
\usepackage{amsmath,amssymb,amsfonts} 
\usepackage[ruled,linesnumbered]{algorithm2e}
\usepackage{graphicx}
\usepackage{textcomp}
\usepackage{xcolor}
\usepackage{multirow}

\usepackage{amsmath}
\usepackage{booktabs} % For formal tables
\usepackage[flushleft]{threeparttable}
\usepackage{pgfplots}
\usepackage{array}
\usepackage{fancybox}
\usepackage{todonotes}
\usepackage{tikz}
\usepackage{textcomp}
\usepackage{todonotes}
\usepackage[utf8]{inputenc}

\usepackage{listings}
\usepackage{color}
\usepackage{tabu}
\selectcolormodel{rgb}

\definecolor{cbxgreenA}    {RGB}{230, 248, 219}
\definecolor{cbxgreenB}    {RGB}{190, 227, 169}
\definecolor{cbxgreenC}    {RGB}{ 88, 171,  30}
\definecolor{cbxgreenD}    {RGB}{ 42,  76,  19}
\definecolor{cbxbluegreen} {RGB}{ 19,  76,  51}
\definecolor{cbxblueA}     {RGB}{223, 238, 255}
\definecolor{cbxblueB}     {RGB}{183, 206, 233}
\definecolor{cbxblueC}     {RGB}{ 69, 136, 214}
\definecolor{cbxblueD}     {RGB}{ 20,  50,  85}
\definecolor{cbxbrownA}    {RGB}{247, 243, 204}
\definecolor{cbxbrownB}    {RGB}{213, 205, 124}
\definecolor{cbxbrownC}    {RGB}{143, 132,  19}
\definecolor{cbxbrownD}    {RGB}{ 84,  67,   0}
\definecolor{cbxredA}      {RGB}{254, 217, 217}
\definecolor{cbxredB}      {RGB}{223, 167, 159}
\definecolor{cbxredC}      {RGB}{176,  24,  24}
\definecolor{cbxredD}      {RGB}{ 98,   9,   9}
\definecolor{cbxblueitem}  {RGB}{ 40, 100, 150}

\colorlet{cbxschemeA}{cbxblueA}
\colorlet{cbxschemeB}{cbxblueB}
\colorlet{cbxschemeC}{cbxblueC}
\colorlet{cbxschemeD}{cbxblueD}

\definecolor{dmlgreen}    {RGB}{51,  160,  44}
\definecolor{dmlblue}     {RGB}{31,  120, 180}
\definecolor{dmlred}      {RGB}{202,   0,  32}

\lstdefinestyle{simple}{%
  language=C++,
  numbers={none},
  basicstyle={\ttfamily},
  moredelim={[is][\underbar]{__}{__}}
}

\definecolor{codegreen}{rgb}{0,0.6,0}
\definecolor{codegray}{rgb}{0.5,0.5,0.5}
\definecolor{codepurple}{rgb}{0.58,0,0.82}
\definecolor{backcolour}{rgb}{0.95,0.95,0.92}

\lstdefinestyle{mystyle}{
	backgroundcolor=\color{white},
	keywordstyle=\color{codegreen},
	numberstyle=\tiny\color{codegray},
	stringstyle=\color{codepurple},
	basicstyle=\ttfamily,
	breakatwhitespace=false,
	breaklines=true,
	captionpos=b,
	keepspaces=true,
	numbers=left,
	numbersep=5pt,
	showspaces=false,
	showstringspaces=false,
	showtabs=false,
	tabsize=2
}

\usepackage[linesnumbered]{algorithm2e}
\usepackage{graphics}

\usepackage{float}
\usepackage{mdwlist}
\usepackage{cases}

\usepackage{multirow}

\usepackage{blindtext}

\usepackage{etoolbox}

\makeatother
\usepackage{enumitem}

\usepackage{gensymb}

\usepackage{tikz}

\usepgfplotslibrary{patchplots}
\usetikzlibrary{pgfplots.statistics}
\usetikzlibrary{snakes}
\usepackage{graphics}
\usepackage{gensymb}
\usepackage{multirow}

\usepackage{xspace}
\usepackage{soul}

\usepackage{amssymb}% http://ctan.org/pkg/amssymb
\usepackage{pifont}% http://ctan.org/pkg/pifont

\usepackage[symbol]{footmisc}

\usepackage{bbding}
\usepackage{makecell}
\usepackage[symbol]{footmisc}

\def\lmtt@use@light@as@normal{}

\newcommand{\cw}{\columnwidth}
\newcommand{\tw}{\textwidth}

\newcommand{\comm}[1]{}

\begin{document}

\title
{%
  ARENA: Asynchronous Reconfigurable Accelerator Ring to Enable Data-Centric \\ Parallel Computing
}

\author{Cheng Tan, Chenhao Xie, Tong Geng, Andres Marquez, Antonino Tumeo, Kevin Barker, and Ang Li%~\IEEEmembership{Member,~IEEE}% <-this % stops a space
\IEEEcompsocitemizethanks{\IEEEcompsocthanksitem The authors are with Pacific Northwest National Laboratory, Richland,
WA, 99354.\protect\\
E-mail: cheng.tan@pnnl.gov}% <-this % stops an unwanted space
}%\thanks{Manuscript received April 19, 2005; revised August 26, 2015.}}

\IEEEtitleabstractindextext{%
\begin{abstract}
The next generation HPC and data centers are likely to be reconfigurable and data-centric due to the trend of hardware specialization and the emergence of data-driven applications. In this paper, we propose ARENA -- an asynchronous reconfigurable accelerator ring architecture as a potential scenario on how the future HPC and data centers will be like. Despite using the coarse-grained reconfigurable arrays (CGRAs) as the substrate platform, our key contribution is not only the CGRA-cluster design itself, but also the ensemble of a new architecture and programming model that enables asynchronous tasking across a cluster of reconfigurable nodes, so as to bring specialized computation to the data rather than the reverse. We presume distributed data storage without asserting any prior knowledge on the data distribution. Hardware specialization occurs at runtime when a task finds the majority of data it requires are available at the present node. In other words, we dynamically generate specialized CGRA accelerators where the data reside. The asynchronous tasking for bringing computation to data is achieved by circulating the task token, which describes the dataflow graphs to be executed for a task, among the CGRA cluster connected by a fast ring network. Evaluations on a set of HPC and data-driven applications across different domains show that ARENA can provide better parallel scalability with reduced data movement (53.9\%). Compared with contemporary compute-centric parallel models, ARENA can bring on average 4.37$\times$ speedup. The synthesized CGRAs and their task-dispatchers only occupy 2.93mm$^2$ chip area under 45nm process technology and can run at 800MHz with on average 759.8mW power consumption. ARENA also supports the concurrent execution of multi-applications, offering ideal architectural support for future high-performance parallel computing and data analytics systems.
\end{abstract}

\begin{IEEEkeywords}
Compute-Flow-Architecture, Runtime Reconfiguration, Asynchronous Parallel Execution, Abstract Machine Model.
\end{IEEEkeywords}}

\maketitle

%\maketitle
\date{}
\pagestyle{plain}
%-------------------------------------------------------------------------
% Front Matter
%-------------------------------------------------------------------------

% \papernum{755}
% \confabbr{CONF XXXX}

\maketitle

%-------------------------------------------------------------------------
% Body
%-------------------------------------------------------------------------
%\input{sec-abstract}
%=========================================================================
% Introduction
%=========================================================================

\section{Introduction}\label{sec:intro}
\vspace{-0.1cm}
\sloppy

With the slowing down of Moore's Law \cite{moore1965cramming}, future computer systems will need to resort to domain-specific accelerators for continuous performance scaling \cite{shalf2020future, dongarra2020numerical} under the same power envelope. This is especially the case for HPC and data centers, as we are quickly entering an era of extreme heterogeneity \cite{vetter2018extreme}, characterized by cluster nodes integrating a multitude of cooperating accelerators \cite{putnam2014reconfigurable, hazelwood2018applied, jouppi2017datacenter, lee2017introducing}.

While integrating domain-specific accelerators (DSAs) into HPC and data centers provides considerable efficiency gains \cite{hazelwood2018applied, jouppi2017datacenter, shaw2014anton, ohmura2014mdgrape}, it leads to enormous complexities as well. First, DSAs can only be economically designed for ubiquitous computational patterns in applications, while the workloads currently running in HPC and data centers are converging towards mixed workflows that include scientific simulation, machine learning, data analytics, etc. Second, managing various (typically loosely coupled) accelerators across many nodes significantly complicates programming models and the software infrastructure. In both HPC and data centers, the accelerators typically need to be shared among multiple users, often using multiple nodes for applications with divergent characteristics. Since we could not design hardware accelerators for all seen and unseen kernels, reconfigurable architecture, which allows specialization after system deployment and even during system execution, promises to be a wise solution. While Field Programmable Gate Arrays (FPGAs) have already been deployed at large scale in some data centers \cite{putnam2014reconfigurable}, they may suffer from limited frequency and energy-efficiency compared with ASICs, as well as long reconfiguration time (e.g. in milliseconds) due to bit-level reconfigurability. Coarse-grained reconfigurable arrays (CGRAs), which integrate highly optimized functional units (rather than fundamental lookup tables or LUTs) and offering reconfigurability at the word-level, emerge as a promising alternative choice \cite{prabhakar2017plasticine}. The rapid reconfiguration \cite{de2019coarse} makes dynamic formalization of hardware accelerators at runtime becomes feasible, and even plausible.

Conventional large-scale HPC clusters implicitly assuming homogeneous node-configuration and bulk-synchronous-parallel (BSP) execution model suffer from three low-utilization challenges regarding per-node data locality: (1) unbalanced data distribution among homogeneous nodes may lead to unbalanced workload and poor utilization; (2) If data is not locally available, during the long-time remote data fetching, the compute units can be idle; (2) Even worse, these idle units or idle nodes cannot be reclaimed for other tasks despite the node may hold their desired data. As emerging workloads become more dynamic and data-driven, decentralized asynchronous task management is highly desired, while data locality becomes a crucial factor for the system design \cite{boroumand2018google, shalf2020future, dongarra2020numerical}. This is largely due to the observation that the energy cost of data-movement significantly overweights the energy cost of computing them \cite{horowitz20141, kestor2013quantifying, molka2010characterizing}, which is particularly the case when migrating data through the interconnect network (e.g., the power budget is $\sim$5.5 watts per full bi-directional NVLink port \cite{nvlink_power}). While software solutions such as \emph{active messages} \cite{bonachea2017gasnet, bonachea2018gasnet} and remote procedure call (RPC) mitigate the problem by pushing computation to the nodes where data resides, they suffer from considerable overhead due to the lack of architectural support. Existing applications however still widely adopt the BSP model \cite{valiant1990bridging, bisseling1993scientific} alternating phases of (parallel) local computation with phases of global communication. Note that the BSP model implicitly assumes that the majority of the time can be spent in easily parallelizable computation phases, with limited data movement. However, the exponential growth in the availability of data and the emergence of new applications, radically changed the balance.

% Existing HPC clusters are largely managed in a centralized fashion, where the login node distributes parallel tasks among computing nodes in a bulk-synchronous fashion. The underlying assumption is that the costs and compute capability per node are homogeneous and ideally constant. Such a conventional compute-centric execution model faces two utilization challenges:

\begin{figure}[htb!]
	\centering
	\includegraphics[width=0.8\cw]{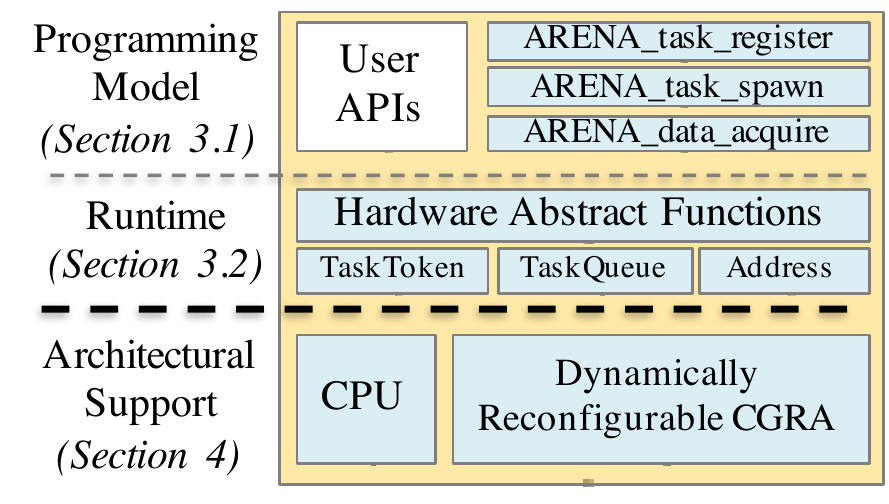}
	\caption{Overall ARENA design stack.}
	\label{fig:stack}
\end{figure}

Bridging coarse-grained reconfigurable architecture with locality-oriented asynchronous parallel task execution, in this paper, we propose ARENA --- an \underline{A}synchronous \underline{RE}co\underline{N}figurable \underline{A}ccelerator ring architecture \& runtime, to enable data-centric computation flow paradigm aiming at significantly reducing unnecessary inter-node data movement. ARENA comprises multiple CGRAs interconnected in a ring to provide quick reconfiguration at runtime. The tasks of the application, tagged as task tokens, are injected into the ring when dependencies are satisfied. Task-specialized CGRA accelerators can be constructed at runtime using the specification embedded in the task token. When task tokens are streaming around the ring, each node can verify whether a task should be executed locally (based on data locality and resource availability). In other words, ARENA is dynamically sharing hardware resource according to data locality. This is in contrast with conventional systems sharing time-slots according to hardware availability. In summary, this paper thus makes the following contributions:

%As task tokens circulating around the ring, each node verifies if the desired data for the present task is locally available, and whether sufficient hardware resources are remaining for initiating the tasks.

%In other words, ARENA is spatially managing the CGRA chip area, which is divergent from current HPC/data-center practice that temporarily allocate time-slots among tasks or user jobs. Meanwhile, this is also the first work enabling asynchronous execution over CGRA clusters, to the best of our knowledge. This paper thus makes the following contributions:
%ARENA aims to investigate a potential design for a scalable multi-node CGRA cluster, introducing solutions to move computation to the data while at the same time providing a reconfigurable substrate that can accelerate different applications. ARENA provides higher efficiency through dynamic hardware specialization and reduction of data movement. 
\begin{itemize}
    \item \textbf{Architecture:} we proposed a multi-CGRA cluster architecture (Section~\ref{sec:architecture}) that can be dynamically reconfigured at runtime, enabling asynchronous parallel execution for multi-users with very little overhead.
    \item \textbf{Runtime:} we proposed a flexible runtime (Section~\ref{subsec:runtime}) enabling task-token delivery along the ring network based on data locality. Compared to compute-centric execution models, the proposed runtime can significantly mitigate data movement and improve performance. 
    \item \textbf{Programming Model:} we proposed a data-centric programming model (Section~\ref{subsec:programming}) with easy-to-use APIs to facilitate the programming for ARENA architecture and runtime. An LLVM-based compiler toolchain is constructed to support the programming model.
    \item \textbf{Evaluation:} RTL simulation results using practical HPC and data analytics applications (Section~\ref{sec:evaluation}) show that ARENA can provide better parallel scalability with reduced data movement (53.9\%) and bring 2.17$\times$ and 4.37$\times$ speedup over traditional compute-centric approach with and without CGRA acceleration. The CGRA-based ARENA prototype only costs 2.93$mm^2$ chip area in 45nm and can work at 800MHz with 759.8mW power consumption per node, showing significant advantages over the present architectural design in the current HPC and data centers. 
\end{itemize}

\section{Motivation}\label{sec:background}
\sloppy

\begin{figure*}[htb]
	\centering
	\includegraphics[width=0.9\tw]{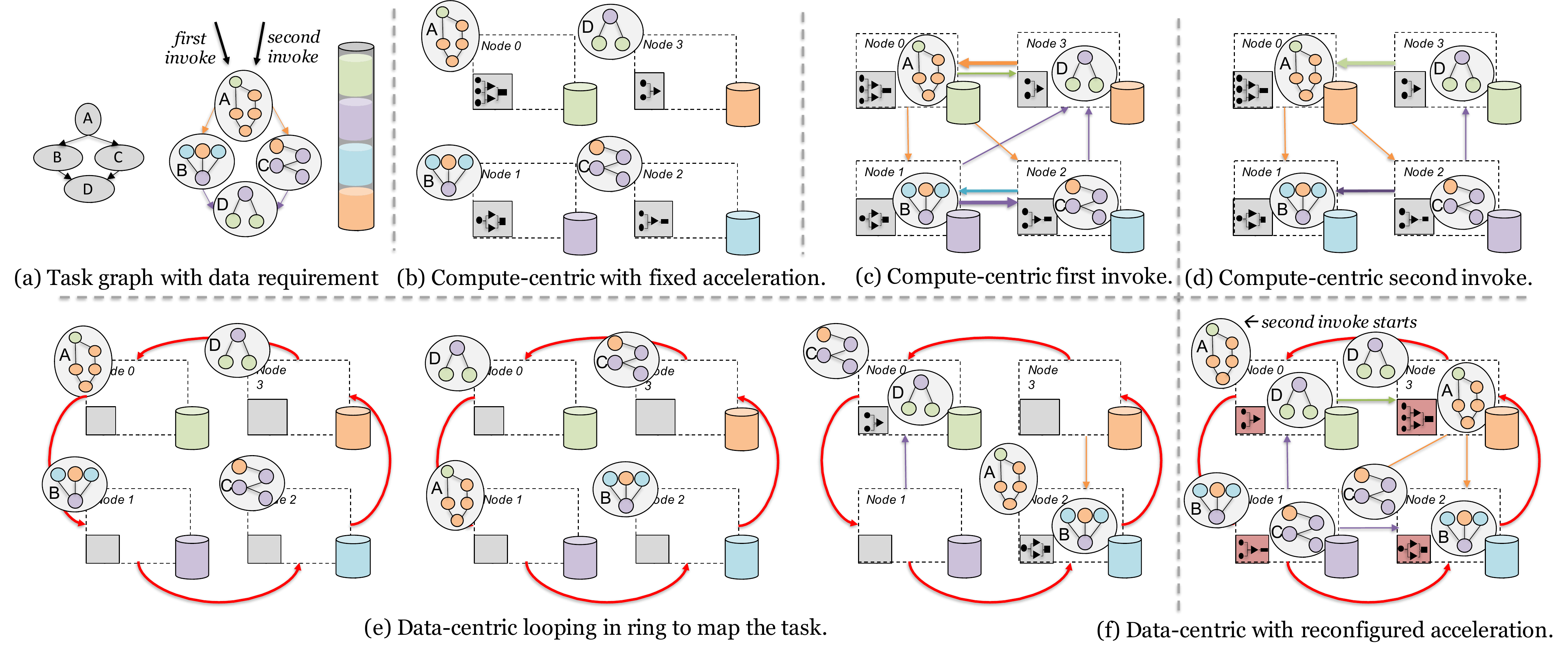}
	\caption{Motivating example comparing traditional compute-centric execution model and ARENA data-centric execution model.}
	\label{fig:motivating}
\end{figure*}

Scientific simulation, where linear solvers iterate on data organized in dense (structured sparse) matrices or tensors, typically easy to divide in equally sized tiles, represents the premier HPC workload \cite{NWChem, GROMACS, quantummech, ExaCT, Carbon}. However, emerging HPC applications, targeting areas such as power grid dynamics~\cite{gridlab}, seismic risk assessment~\cite{Seismic}, urban systems simulation, and microbiome analysis~\cite{Genome}, will likely combine together traditional scientific simulation with advanced data analytics and machine learning. The datasets for these applications are much less structured, and thus more difficult to organize in \textbf{regular} and \textbf{partitionable} data structures. Applications will alternate phases of scientific simulations with \textbf{regular} behaviors, to phases where the computation happens on sparse data structures (e.g., sparse matrices, graph traversal) that induce unpredictable fine-grained data accesses and \textbf{irregular} behaviors. This scenario provides a clear opportunity for adaptability to diverse behaviors with reconfigurable hardware, while at the same time makes the current HPC programming models inadequate. In the following, we describe three major existing multi-node computing paradigms to motivate the ARENA design. 

\subsection{Baseline-1: Compute-Centric BSP Execution Model}

HPC systems typically rely on the classical Bulk Synchronous Parallel (BSP) programming model in which a process is assigned to a processor or an entire node, and communication typically happens through message passing with libraries such as MPI. In the BSP model, the computation proceeds as a series of global supersteps: concurrent computation, where every participating nodes perform parallel computations on local data; communication, where nodes exchange data among them (with various, algorithm dependent, patterns); and barrier synchronization, to align execution of nodes. 

The BSP model assumes that data are partitioned and distributed across nodes and rarely moves, to facilitate the local computation super-steps. Otherwise, the message-passing based communication super-steps would dominant the execution time. While this model works well for applications with easily partitionable data, regular computation, and limited, structured communication, it starts to experience significant limitations when workloads exhibit irregular behaviors (skewed data distributions, high synchronization intensity, irregular communication patterns). For these reasons, we consider the BSP model as \emph{Compute-Centric}.

Consider as an example an application with the (hierarchical) task graph in Figure~\ref{fig:motivating}(a), where the computation is split into 4 high-level tasks, each one executing task-partitioned computational kernels where the subtasks require data from other nodes. When employing a BSP model, both the data allocation and the distribution of the high-level tasks are fixed for the entire application execution. Hence, if a subtask needs data available in another node, it needs to initiate a communication phase, load the remote data, and synchronize to avoid hazards. Despite the latest high-performance designs can exploit mechanisms such as \emph{remote-direct-memory-access} (one-sided communication, not requiring a blocking receive with implicit synchronization from the remote node), prefetching and data migration, when these remote accessing are frequent, the bandwidth gap between local memory and remote memory (which needs to be accessed through the network) still remains the major performance (and consequently energy) concern. For example, if, as typical in HPC applications, tasks are executed in a loop, and they contend on the same data blocks (e.g., Tasks A, B, and C in Figure \ref{fig:motivating}), data migration may trigger even more data movement and synchronization, as the actual data distribution and access patterns are unknown before runtime.

\textbf{Limitations --} While compute-centric BSP long remains the standard model in HPC, its adoption in emerging HPC applications may be limited by data movement and synchronization. HPC practitioners have introduced asynchronous multi-task runtimes to tackle the limitations of the BSP model. Besides migrating data blocks where the computation occurs, these runtimes often allow tasks to migrate where the data reside, via approaches such as active messages and remote procedure calls, following the \emph{data-centric} models.

\subsection{Baseline-2: Reconfigurable accelerators in HPC}

Reconfigurable accelerators have been deployed at a massive scale in data centers to provide application-specific acceleration with improved power-/area-efficiency~\cite{putnam2014reconfigurable, prabhakar2017plasticine, geng2018fpdeep}. Some institutions have also started hosting clusters with FPGA to perform research in HPC \cite{Paderborn}. 

However, accelerators in HPC installations still generally adopt the BSP model, where a part of the local computation is offloaded to the accelerator itself. This also requires gathering desired data by the running tasks offloaded for execution. While FPGAs potentially allow acceleration of workloads more diverse than conventional accelerators such as GPUs, their current usage in HPC application often entails static configuration for accelerating a small amount of kernels (e.g., FFT, GEMM). This approach matches with an offload accelerator model - tasks and computational kernels do not move in the system. On the other hand, a reconfigurable architecture potentially allows to dynamically adjust the configuration at runtime, being able to accelerate divergent tasks with the computation proceeds. Despite the potential, the excessive reconfiguration overheads (typically in milliseconds) makes such an approach very costly.

%\subsection{Baseline-2: Data-Flow Execution Model}

%Reconfigurable accelerators have been deployed at massive scale in data centers to provide application-specifc acceleration with improved power-/area-efficiency~\cite{putnam2014reconfigurable, prabhakar2017plasticine, geng2018fpdeep}. Synchronizations are alleviated and only happened at the dependent kernels. In conventional compute-centric systems, if an accelerator is provided on a node, it is always configured to maximize the speedup of the target computation (as shown in Figure \ref{fig:motivating}(c)-(d)). For example, typically, an FPGA board is specialized to a single application kernel during runtime. However, if the compute workload on a node changes, the accelerator should be dynamically reconfigured, which would take as long as milliseconds for reconfiguration on the FPGA.

%\textbf{Limitations --} Data-Flow hardware alleviates the synchronization overhead and provides high utilization hardware acceleration for a fixed task on one node but lack of quick runtime support for multiple different tasks, which limits the overall throughput. In addition, the data-flow streamed between tasks incurs even more data movement.
\textbf{Limitations --} Reconfigurable accelerators make it possible to accelerate a more diverse workload, but current practice in heterogeneous HPC still leverages the offload model. The lack of low-latency runtime reconfiguration also limits the chances of large-scale task migration across the whole system. On the other hand, the data-centric execution model allows different tasks to work concurrently, despite using the same set of data in a node. CGRAs, as an alternative reconfigurable solution \cite{prabhakar2017plasticine} to FPGAs~\cite{de2019coarse}, offer significantly reduced reconfiguration time with coarser-grained reconfiguration, making flexible runtime architecture adjustment becomes possible.

%CGRA are two typical reconfigurable architectures. FPGAs provide higher reconfigurability but with lower power and area efficiency due to the bit-level reconfiguration overhead. On the other hand, CGRA enables word-level reconfigurability with less reconfiguration overhead leading to better efficiency than FPGA~\cite{de2019coarse}.

%In addition, the maximum acceleration speedup is theoretically not only limited by the available accelerator resources, but also data parallelism inside the computation (as shown in Figure~\ref{fig:cgra_speedup}). Therefore, monopolizing the entire accelerator on a node for single task bottlenecks the overall throughput and decrease area-efficiency. To solve this problem, ARENA dynamically reconfigures the accelerator for the present task based on its data requirement and the current available accelerator resources.

%Specifically, the computation fetches the data in, performs computation with maximized acceleration, and stores the results locally or routes the data out towards the destination node. Computation is invoked in order while data is streaming. 

\subsection{Baseline-3: Data-Centric Software Approaches}

Many programming models have been leveraged or designed to allow data-centric execution on multi-node systems. SSMP~\cite{perez2010handling} can operate on shared memory machines and support dynamic detection of dependencies between tasks. The implicitly shared memory management and dependence detection improve the programmability at the cost of increased synchronization overhead. \emph{Remote procedure call} (RPC) achieves near-data-computation based on prior knowledge about the exact distribution of data. X10~\cite{charles2005x10} and Chapel~\cite{chamberlain2007parallel} allow the programmers to control where to place the data and tasks. Similarly, Legion~\cite{bauer2012legion} enables explicit, programmer-controlled movement of data and placement of asynchronously spawned tasks, based on locality information. Legion employs a Cilk-like~\cite{blumofe1996cilk} algorithm for locality-aware task stealing. 

Towards data-centric programming, MapReduce~\cite{dean2008mapreduce} programmers think in a data-centric fashion: they focus more on handling the sets of data records, rather than managing fine-grained threads, processes, communication, and coordination~\cite{alvaro2009boom}. However, MapReduce constrains its usage to batch-processing tasks, which falls in the BSP scope. To accommodate the emerging data-driven applications with irregular and unpredictable data access patterns, data-centric execution with asynchronous task-spawn should be enabled with hardware support. Meanwhile, the synchronization and task dependencies should be specified by the programmers to eliminate the unnecessary performance and energy overhead rather than forced by the programming model (e.g., remote procedure need to return to local in RPC, all the spawned tasks of the same ancestor need to join eventually in Legion). Besides the data locality, the runtime should also consider the computing resource utilization when reconfigurable accelerators are deployed and shared by multi-users in HPC/data-center environments.

\textbf{Limitations --} The high-level software framework and runtime facilitate the asynchronous execution of tasks and attempt to take advantage of data locality. Unfortunately, existing frameworks based on software solutions incur considerable overhead. The lack of hardware reconfiguration also limits the benefit from application-specific design and heterogeneity.

\subsection{Compute-Flow ARENA Execution Model}

We propose ARENA to address the limitation of the three aforementioned baseline. ARENA includes a novel programming model targeting asynchronous data-centric execution paradigm. As shown in Figure~\ref{fig:motivating}, all the configurable nodes in ARENA are connected by a ring network to bring the specialized computation to the data rather than the reverse to minimize data movement. Each reconfigurable node mainly contains a CGRA (detailed in Section~\ref{subsec:cgra}) that supports real-time reconfiguration and simultaneous execution of multi-tasks.

%\begin{figure}[htb]
%	\centering
%	\includegraphics[width=\cw]{graphs/category.pdf}
%	\vspace{-0.3cm}
%	\caption{Programming model and runtime.}
%	\label{fig:Non_essential}
%\end{figure}

%=========================================================================
% Programming
%=========================================================================

\section{ARENA Programming Model}\label{sec:programming}

Being the interface between software and hardware, the ARENA programming model defines a list of API functions in Table~\ref{tab:program_interface}. On one hand, in order to program an ARENA abstract machine, a software programmer has to define their user-logic as task functions, and rely on the \emph{User APIs} to operate the abstract machine. Please note that although in this work we use CGRAs as the hardware testbed, it is only one of the the possible instantiations of the ARENA abstract machine model (AMM). On the other hand, in order to support ARENA software and run ARENA program, an alternative architecture has to support the \emph{Hardware Abstract Functions} here.

\begin{table}[htb!]
\scriptsize
\centering
\begin{tabular}{l|l}
\bottomrule\bottomrule
\textbf{Function \& Construct} & \textbf{Description} \\ \toprule
\multicolumn{2}{c}{\textbf{User Defined Function}} \\ \bottomrule
\begin{tabular}[c]{@{}l@{}}void \textbf{my\_task}(\\\hspace{0.7cm}\textbf{Address} \textit{TASKstart},\\\hspace{0.7cm}\textbf{Address} \textit{TASKend},\\\hspace{0.7cm}\textbf{float} \textit{PARAM})\end{tabular}& \begin{tabular}[c]{@{}l@{}}A user can implement multiple different \\tasks to compose a single or multiple \\applications. \end{tabular}\\ \toprule
%\begin{tabular}[c]{@{}l@{}}void \textbf{recv\_data}(\\\hspace{0.7cm}\textbf{Address} \textit{RECVstart},\\\hspace{0.7cm}\textbf{Address} \textit{RECVend}) \end{tabular}& \begin{tabular}[c]{@{}l@{}}Acquires additional data \\from remote node.\end{tabular} \\ \toprule
\multicolumn{2}{c}{\textbf{User APIs}} \\ \bottomrule
\begin{tabular}[c]{@{}l@{}}void \textbf{ARENA\_task\_register}(\\\hspace{0.7cm}\textbf{int} \textit{TASKid},\\\hspace{0.7cm}\textbf{Address} \&\textit{my\_kernel},\\\hspace{0.7cm}\textbf{bool} \textit{isRoot})\end{tabular}& \begin{tabular}[c]{@{}l@{}} Registers a kernel (e.g. \textit{my\_kernel}) with\\ \textit{TASKid}. The root task is launched by a \\CPU or a microcontroller once the system \\starts to run. \end{tabular}\\ \hline
\begin{tabular}[c]{@{}l@{}}void \textbf{ARENA\_task\_spawn}(\\\hspace{0.7cm}\textbf{int} \textit{TASKid},\\\hspace{0.7cm}\textbf{Address} \textit{TASKstart},\\\hspace{0.7cm}\textbf{Address} \textit{TASKend}, \\\hspace{0.7cm}\textbf{float} \textit{PARAM}, \\\hspace{0.7cm}\textbf{Address} \textit{REMOTEstart}, \\\hspace{0.7cm}\textbf{Address} \textit{REMOTEend})\end{tabular}& \begin{tabular}[c]{@{}l@{}}Dynamically spawns a new \textbf{TaskToken} \\(\textit{FROMnode} is automatically applied) that \\will be issued to the \textbf{CoalesceUnit}. The \\fields of a task token is explained in detail \\in Section~\ref{subsec:token}. \end{tabular}\\  \toprule
\multicolumn{2}{c}{\textbf{Hardware Abstract Functions}} \\ \bottomrule
\begin{tabular}[c]{@{}l@{}}void \textbf{ARENA\_init}(\\\hspace{0.7cm}\textbf{Address*} \textit{local\_start},\\\hspace{0.7cm}\textbf{Address*} \textit{local\_end})\end{tabular}& \begin{tabular}[c]{@{}l@{}}Initializes \textit{local\_start} and \textit{local\_end} based \\on local data information. \end{tabular}\\ \hline
\begin{tabular}[c]{@{}l@{}}{TaskToken} \textbf{ARENA\_arrive}()\end{tabular}& \begin{tabular}[c]{@{}l@{}}Receives an incoming task from a remote \\node.\end{tabular} \\ \hline
\begin{tabular}[c]{@{}l@{}}void \textbf{ARENA\_filter}(\\\hspace{0.7cm}\textbf{TaskToken} \textit{token},\\\hspace{0.7cm}\textbf{Address*} \textit{local\_start},\\\hspace{0.7cm}\textbf{Address*} \textit{local\_end},\\\hspace{0.7cm}\textbf{TaskQueue*} \textit{SendQueue},\\\hspace{0.7cm}\textbf{TaskQueue*} \textit{WaitQueue})\end{tabular}& \begin{tabular}[c]{@{}l@{}}Detaches, Splits or passes tasks based on \\the \textit{token}'s required data addresses (by \\comparing them with \textit{local\_start} and \\\textit{local\_end}).\end{tabular} \\ \hline
\begin{tabular}[c]{@{}l@{}}bool \textbf{ARENA\_ready}(\\\hspace{0.7cm}\textbf{TaskToken} \textit{token})\end{tabular}& \begin{tabular}[c]{@{}l@{}}Checks if the computing resources are \\available for executing the task \textit{token}.\end{tabular} \\ \hline
\begin{tabular}[c]{@{}l@{}}void \textbf{ARENA\_launch}(\\\hspace{0.7cm}\textbf{TaskToken} \textit{token},\\\hspace{0.7cm}\textbf{TaskQueue*} \textit{CoalesceUnit})\end{tabular}& \begin{tabular}[c]{@{}l@{}}Issues a task denoted by \textit{token} either to a \\CPU or to a reconfigurable accelerator (e.g., \\CGRA). The spawned new tasks will be \\pushed into CoalesceUnit.\end{tabular} \\ \hline
\begin{tabular}[c]{@{}l@{}}void \textbf{ARENA\_data\_acquire}(\\\hspace{0.7cm}\textbf{TaskToken} \textit{token}) \end{tabular}& \begin{tabular}[c]{@{}l@{}}Acquires additional data from remote node \\via NIC.\end{tabular} \\ \hline
\begin{tabular}[c]{@{}l@{}}TaskToken \textbf{ARENA\_coalesce}(\\\hspace{0.7cm}\textbf{TaskQueue*} \textit{CoalesceUnit})\end{tabular}& \begin{tabular}[c]{@{}l@{}}Coalesces spawned tasks with continuous \\data range, identical required remote data \\range, and identical PARAM.\end{tabular} 
\\ \toprule
\multicolumn{2}{c}{\textbf{Base Constructs}} \\ \bottomrule
%\begin{tabular}[c]{@{}l@{}}\textbf{TaskToken}(\\\hspace{0.7cm}\textbf{int} \textit{TASKid},\\\hspace{0.7cm}\textbf{Address} \textit{TASKstart},\\\hspace{0.7cm}\textbf{Address} \textit{TASKend}, \\\hspace{0.7cm}\textbf{float} \textit{PARAM}, \\\hspace{0.7cm}\textbf{Address} \textit{REMOTEstart}, \\\hspace{0.7cm}\textbf{Address} \textit{REMOTEend}, \\\hspace{0.7cm}\textbf{Node} \textit{FROMnode})\end{tabular} & \begin{tabular}[c]{@{}l@{}}A task token including task\\ID, task data range and \\ parameter for computation, \\ start/end addresses for \\additional remote data, \\and the node that spawns\\ this token.\end{tabular} \\ \hline
\begin{tabular}[c]{@{}l@{}}\textbf{TaskToken} \end{tabular} & \begin{tabular}[c]{@{}l@{}}encapsulates a task.\end{tabular} \\ \hline
\begin{tabular}[c]{@{}l@{}}\textbf{Address} \end{tabular} & \begin{tabular}[c]{@{}l@{}}Local address of data.\end{tabular} \\ \hline
\begin{tabular}[c]{@{}l@{}}\textbf{TaskQueue} \end{tabular} & \begin{tabular}[c]{@{}l@{}}Buffers for task tokens.\end{tabular} \\ \toprule\toprule
\end{tabular}
\caption{ARENA programming and hardware APIs.}
\label{tab:program_interface}
\end{table}

\subsection{ARENA Programming Interface}\label{subsec:programming}

To program an ARENA abstract machine, the programmers first partition their application into tasks and register the defined tasks to the ARENA runtime. Ideally, the partition can separate the working set into a bunch of continuous data segments, where each task accounts for a segment. ARENA does not limit the granularity of a task, which can be extremely fine-grained or coarse-grained. While ARENA works perfectly when the data is locally available for a task, we understand this is not always feasible. When remote data access is inevitable, the application can either spawn a new task for the remote data, or explicitly initiate the data-movement through the \emph{data-transfer-network}. 

%We will discuss this in detail later.  

\begin{figure}[htb]
\begin{lstlisting}[basicstyle={\scriptsize\fontencoding{T1}\footnotesize\fontfamily{lmtt}\fontseries{c}\selectfont},escapechar=!]
// Users can define multiple different kernels,
// each with a specific task token ID.
#define BFS_TOKEN 1
int** local_M;
...
void BFS_kernel(int TASKstart, int TASKend,
                int PARAM) {
  int level = PARAM;
  for(int i=TASKstart; i<TASKend; ++i) {
    for(int j=0; j<SIZE; ++j) {
      if(local_M[i][j] > level) {
        local_M[i][j] = level;
        ARENA_task_spawn(BFS_TOKEN, j, j+1,
                         level+1); }}}}
void run() {
  // Register kernels.
  ARENA_task_register(BFS_TOKEN, &BFS_kernel, true);
  // Launch ARENA runtime.
  ARENA_runtime(); }
\end{lstlisting}
\caption{Example of programming SSSP in ARENA.}
\label{fig:bfs_arena_impl}
\end{figure}

%void ARENA_data_acquire(TaskToken token) {
%  /* ... Not necessary for BFS. */ }

Figure~\ref{fig:bfs_arena_impl} shows an example on how to solve the single-source shortest path (SSSP) problem using a breadth-first search (BFS) kernel. The design traverses associated vertices until the shortest path(s) from a source node to all the other nodes are found. Without losing generality, we assume the graph, represented as an adjacency matrix, is distributively stored on all nodes and each node holds $SIZE/NODES$ vertices (rows) of the entire graph (the adjacency matrix is in $SIZE$ x $SIZE$, an initial value of $\infty$ indicates a connected edge while 0 implies no connection).

ARENA enables asynchronous data-centric execution by dynamically spawning new task-tokens among the nodes. Currently, all tasks need to be registered at the beginning. During runtime, task-tokens are circulating among all nodes in the ring. In case a node confirms it has the data required by a task-token (indicated by the starting \& ending addresses $TASK_{start}$ and $TASK_{end}$) as well as sufficient hardware resources for runtime hardware specialization implied by the task-token, it takes out the token from the task-token-stream and executes it. A task will be executed by the CPU if no hardware specialization is provided. New tasks for remote nodes can be generated or spawned at any node. We currently rely on the programmer to determine the granularity of a spawned task (in other words, the data-range a spawned task designated). Fine-grained tasks facilitate asynchronous execution but increase scheduling overhead in the runtime. We discuss this tradeoff in detail in Section~\ref{subsec:runtime}.

This is in contrast with the conventional compute-centric approach demanding frequent data communication and synchronization~\cite{bulucc2011parallel}. As each node maintains the vertex status locally, and no prior knowledge about vertex distribution is asserted, repeated all-to-all communications are essentially desired for broadcasting vertex updating information to associated nodes on the present frontier. Figure~\ref{fig:data_centric_speedup} shows the performance gain of ARENA for the SSSP application.

 %Figure~\ref{fig:data_centric_speedup} shows that the data-centric execution model improves the performance with better scalability than the conventional compute-centric one for this single source shortest path application.

%=========================================================================
% Runtime
%=========================================================================

\subsection{ARENA Runtime Support}\label{subsec:runtime}

\begin{figure}[htb!]
\centering
%\begin{subfigure}
\includegraphics[width=0.9\cw]{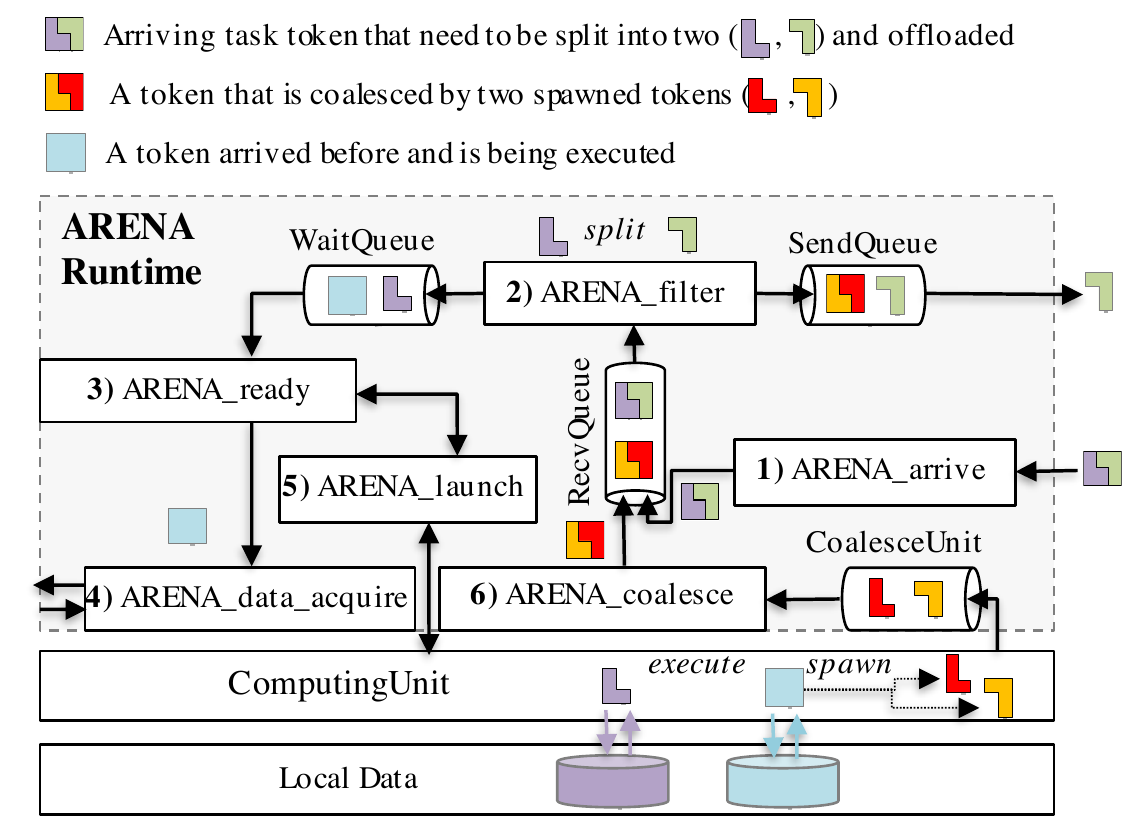}
\caption{ARENA runtime workflow.}
\label{fig:runtime_demo}
\end{figure}

Figure~\ref{fig:runtime_demo} and \ref{fig:runtime_code} illustrate the workflow of ARENA runtime executed per-node and the pseudo-code of the workflow, respectively. As can be seen, multiple tasks (marked with different colors) can be asynchronously executed in parallel. Note, the runtime can be supported by CPUs, GPUs, DSPs, or any other fixed or reconfigurable hardware substrate given the substrate realize the \emph{Hardware Abstract Functions} and support the \emph{Base Constructs}.

\begin{figure}[htb!]
\begin{lstlisting}[basicstyle={\scriptsize\fontencoding{T1}\footnotesize\fontfamily{lmtt}\fontseries{c}\selectfont},escapechar=!]
void ARENA_runtime() {
  bool terminate = false;
  TaskQueue WaitQueue, RecvQueue;!\label{algo:runtime:queue:start}!
  TaskQueue SendQueue, CoalesceUnit;!\label{algo:runtime:queue:end}!
  TaskToken token;
  Address local_start, local_end;
  ARENA_init(&local_start, &local_end);!\label{algo:runtime:init}!
  while(true) {!\label{algo:runtime:while}!
    // Enqueue an arriving task token.
    token = ARENA_arrive();
    RecvQueue.enqueue(token);!\label{algo:runtime:arrive}!
    // Check termination for all tasks.!\label{algo:runtime:terminate:start}!
    token = RecvQueue.dequeue();
    if(token.TASKid == TERMINATE and
       WaitQueue.empty()) {
      SendQueue.enqueue(token);
      if(terminate) break;
      terminate = true;
      continue; }
    terminate = false;!\label{algo:runtime:terminate:end}!
    // Offload, split, or convey a task.!\label{algo:runtime:filter:start}!
    ARENA_filter(token, local_start, local_end,
                 &SendQueue, &WaitQueue);!\label{algo:runtime:filter:end}!
    // Check availability for task execution.
    token = WaitQueue.peek();
    if(ARENA_ready(token)) {!\label{algo:runtime:ready}!
      WaitQueue.dequeue();
      // Acquire data if necessary.
      if(token.REMOTEend > token.REMOTEstart) {
        ARENA_data_acquire(token);}!\label{algo:runtime:acquire}!
      // Issue a task for execution and return
      // without waiting for completion.
      ARENA_launch(token, CoalesceUnit);{!\label{algo:runtime:launch}!
      // Task coalescing if possible.
      token = ARENA_coalesce(CoalesceUnit);{!\label{algo:runtime:coalesce}!
      if(token)
        RecvQueue.enqueue(token); }}}
\end{lstlisting}
%\end{subfigure}
\caption{ARENA runtime pseudocode.}
\label{fig:runtime_code}
\end{figure}

\newcommand{\RNum}[1]{\uppercase\expandafter{\romannumeral #1\relax}}

We describe the runtime process below: Primarily, the task queues (line~\ref{algo:runtime:queue:start}-\ref{algo:runtime:queue:end}) and local data range (line~\ref{algo:runtime:init}) are initialized. We then proceed with 6 steps: \textbf{Step-(1):} All the incoming task tokens from the proceeding node will be appended to the \texttt{RecvQueue} (line~\ref{algo:runtime:while}-\ref{algo:runtime:arrive}). \textbf{Step-(2):} A token popped from the \texttt{RecvQueue} will be processed in the \texttt{Filter}, where a task can be split into multiple tasks, which are either buffered in the \texttt{WaitQueue} for local execution, or forwarded to the \texttt{SendQueue}, wait for being conveyed to the next node (line~\ref{algo:runtime:filter:start}-\ref{algo:runtime:filter:end}). The logic is: (\RNum{1}) if the task data range is irrelevant to the node's local data range, it is forwarded to \texttt{SendQueue} as it is; (\RNum{2}) if the task data range is a subset of the node's local data range ($local_{start} \leq TASK_{start} \leq TASK_{end} \leq local_{end}$), implying the local availability of all the needed data for the task, the token will be pushed into \texttt{WaitQueue} for future execution; (\RNum{3}) if the data range indicated by the task is a superset of the node's local data range ($TASK_{start} \leq local_{start} \leq local_{end} \leq TASK_{end}$), it suggests that the task might be too coarse-grained. We split the task into three portions and spawn three new tasks. The one with data range $local_{start}$ to $local\_{end}$ will be buffered in \texttt{WaitQueue} for local processing; the other two tasks are redirected to \texttt{SendQueue}; (\RNum{4}) finally, if the task data range is partially aligned with the node's local data range, two new tasks will be spawned. The aligned part is buffered in \texttt{WaitQueue} while the mismatch part is forwarded to \texttt{SendQueue}. \textbf{Step-(3):} The runtime checks whether there is available resources for the token at the top of the \texttt{WaitQueue} to be executed (line~\ref{algo:runtime:ready}). \textbf{Step-(4):} If so, the runtime verifies whether the task needs to incur any inevitable remote data access. If yes, the data will be acquired from the objective remote nodes (line~\ref{algo:runtime:acquire}) through the \texttt{Data-Transfer-Network}. \textbf{Step-(5):} when all demanding data are available, the task is issued to the computing resources for execution (line~\ref{algo:runtime:launch}). \textbf{Step-(6)} As new tasks can be generated locally in a node during the task execution, to avoid too many tasks flooding the system, a \texttt{CoalesceUnit} (line~\ref{algo:runtime:coalesce}) is designed to aggregate the newly generated tasks if the boundaries of their data-ranges coincide each other, and whether alternative key parameters ((e.g., task-token carried partial-reduction variables) are the same. This can avoid the scenario that too many generated fine-grained tasks saturate the task-token network and the associated buffers. Finally, the runtime on a particular node terminates when the TERMINATE token has been continuously received, and there is no pending tasks in the local \texttt{WaitQueue} (line~\ref{algo:runtime:terminate:start}-\ref{algo:runtime:terminate:end}).

\section{ARENA CGRA-Cluster Architecture}\label{sec:architecture}
\sloppy

\begin{figure}[htb!]
	\centering
	\includegraphics[width=0.99\cw]{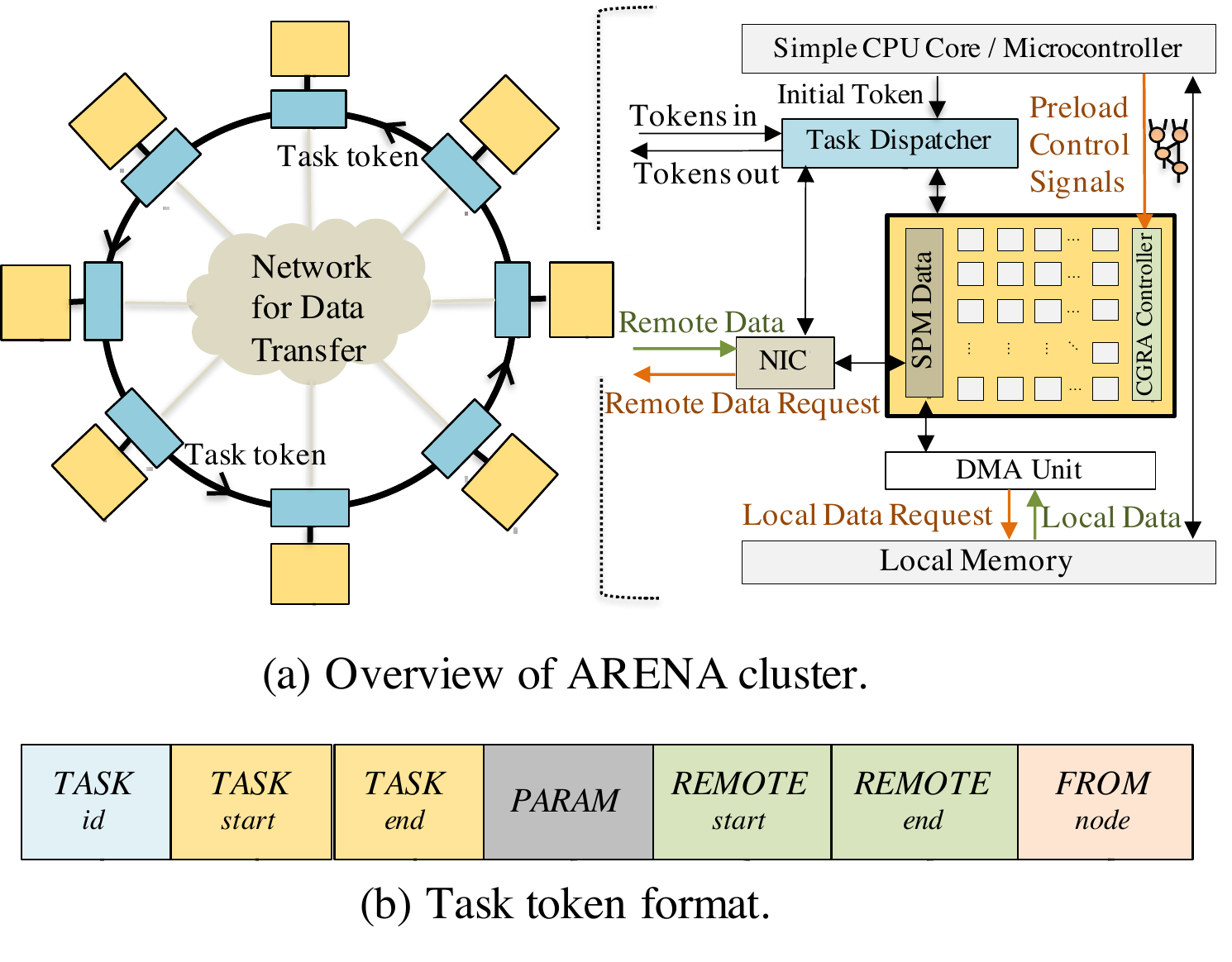}
	\caption{ARENA cluster overview and task token format.}
	\label{fig:framework}
\end{figure}

Our ARENA prototype in this work is built upon a CGRA cluster interconnected by a fast ring network. Figure~\ref{fig:framework}(a) shows the overall design --- multiple reconfigurable nodes are connected in a ring topology. Each node incorporates a micro-controller (e.g., a simplified CPU), a task dispatcher, a CGRA, a network interface (NIC), a DMA unit, and local memory storage. At runtime, task tokens are circulating along the ring. The dispatcher can split, offload, and forward a task token based on the data range as already discussed. We adopt the ring network topology to simplify the routing strategy and provide sufficient bandwidth for delivering the small-sized task tokens ($\sim$21 bytes) with up to 16 nodes evaluated in this paper. The ring network can also provide near optimal bandwidth for most collective communication \cite{nvidia2017dgx1}, and can be easily built upon various physical network topology. We leave the exploration of alternative multi-node topology as a future work. 

% Figure 6 illustrates the block diagram for the system with the 4 × 4 UE-CGRA. To offload computation, the processor writes the control status registers (CSRs) in the CGRA through the accelerator command interface and sets up the base addresses and sizes for the DMA unit to fetch the config- uration bitstream and the data required for the computation. The data is loaded into the SRAM banks inside the PE ar- ray. Another CSR is then written to initiate computation. The complete UE-CGRA is composed of a control unit, a DMA unit with read and write queues, and an array of UE-CGRA PEs interconnected with queues. The PEs along the north and south perimeter contain SRAM banks and are the only PEs capable of memory operations. The PE is carefully ar- chitected to enable both compute and bypassing of data (i.e., routing) in the same cycle. The configuration phase leverages the existing data network to forward configuration messages systolically through the array from top to bottom. Each block is described in more detail below.

\subsection{Task Token}\label{subsec:token}

A task is represented by a task token in ARENA, which can be dynamically spawned, executed, delivered, and split. Figure~\ref{fig:framework}(b) shows the general format of the task token which comprises 7 fields: $TASK_{id}$ indicates the task will be executed, which is registered by the user (using \emph{ARENA\_task\_register()}) before launching the ARENA runtime. During execution, the reconfigurable node will be dynamically configured (see Section~\ref{subsec:cgra}) based on $TASK_{id}$. $TASK_{start}$ and $TASK_{end}$ together describe the data range for the task. $PARAM$ refers to a token-carried return value that is typically initialized by its parent task. This field is useful when performing collective operations (e.g., \emph{reduction} and \emph{accumulation}). For unavoidable remote data access, we use $REMOTE_{start}$ and $REMOTE_{end}$ to indicate the starting and ending addresses. $FROM_{node}$ labels the node where its parent task locates. Each task token thus requires 21 bytes in our prototype architecture (i.e., 4-bit each for $TASK_{id}$ and $FROM_{node}$; 4-byte each for the other fields).

%ARENA's CGRA supports $spawn$ operation that fills all the fields of the task token to spawn a new task and issue it into the CGRA controller. 

\subsection{Task Dispatcher}

\begin{figure}[htb]
	\centering
	\includegraphics[width=0.9\cw]{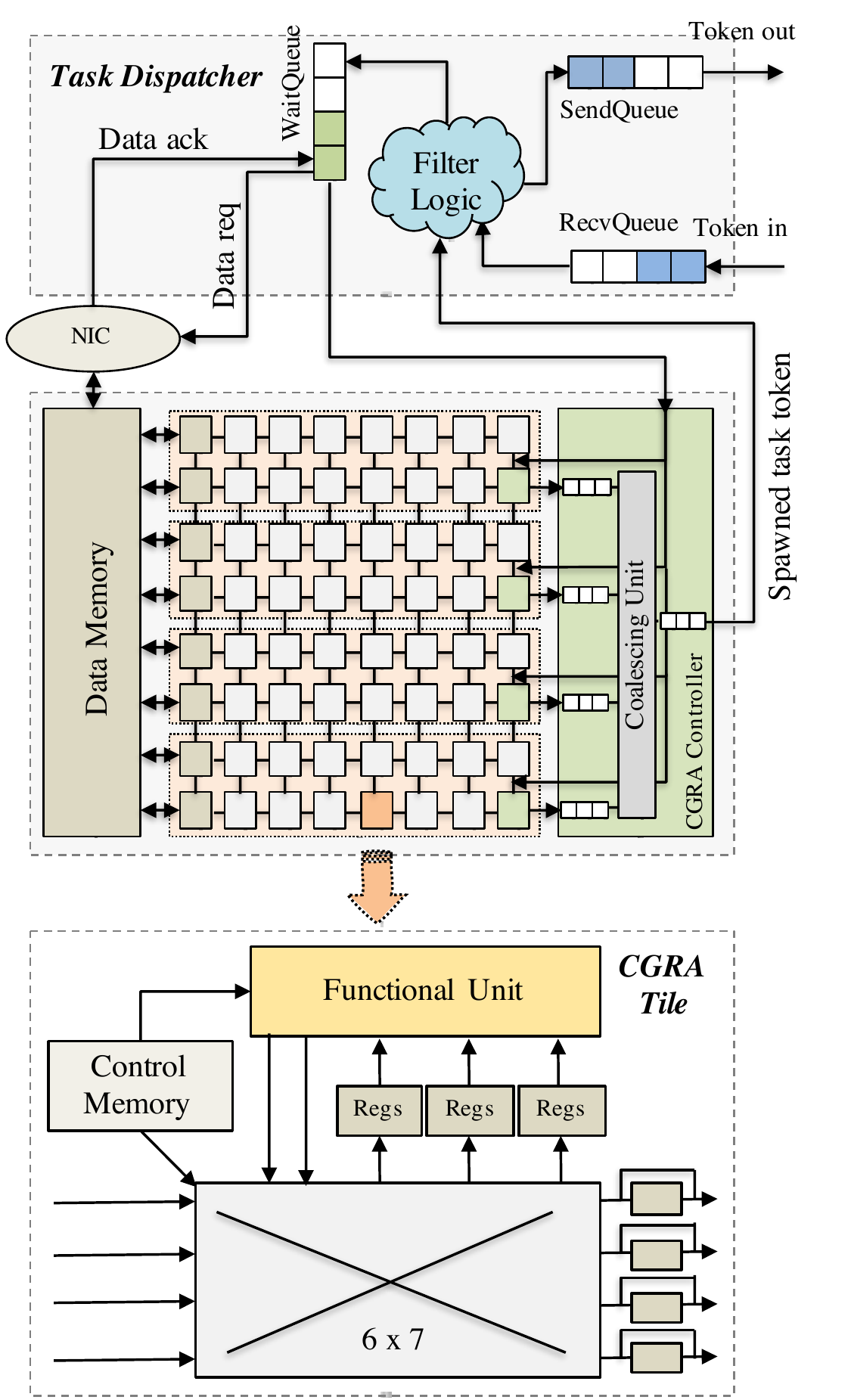}
	\caption{{Architecture of an ARENA node.}}
	\label{fig:arena_node}
\end{figure}

ARENA's task dispatcher mainly includes a task \texttt{Filter Logic} and three queues as shown in Figure~\ref{fig:arena_node}. The \texttt{Filter Logic} can offload, split, and convey a task token based on the data requirement (see Section \ref{subsec:runtime}). The NIC handles remote data requests from the task tokens in the \texttt{WaitQueue}. The \texttt{WaitQueue} will be acknowledged when the required remote data arrives at the data memory. The CGRA controller will then pop the acknowledged task token from the head of \texttt{WaitQueue} and reconfigure the CGRAs accordingly.

%A task token is first pushed into a \textbf{Receive Queue} when it arrives at the dispatcher and the receive queue is not full. The \textbf{Filter Logic} pops the task token from the top of the receive queue and analyzes it. Specifically, if the task data range embedded in the task token is a subset of the local data range (indicated by $local\_start$ and $local\_end$, as every node is only aware of its own local data range), i.e., $TASKstart \geq local\_start$ and $TASKend \leq local\_end$, the task token will be pushed into the \textbf{Wait Queue}. If the task data range is exclusive to the local data range (i.e., $TASKend$ \leq $local\_start$ or $TASKstart$ \geq $local\_end$), the task tag will be pushed into the \textbf{Send Queue}. If the task data range is a superset of the local data range (i.e., $TASK_{start} < LOCAL_{start}$ and $TASK_{end} > local\_end$), the filter logic will split the original task tag into three new ones (i.e., $task_a$ with task data range from $TASK_{start}$ to $LOCAL_{start}$, $task_b$ with task data range from $LOCAL_{start}$ to $LOCAL_{end}$, and $task_c$ with task data range from $LOCAL_{end}$ to $TASK_{end}$). Both $task_a$ and $task_c$ will be pushed into the send queue while $task_b$ will be pushed into the wait queue. 

\subsection{Reconfigurable CGRA Nodes and Toolchain}\label{subsec:cgra}

%FPGAs and CGRAs are two typical reconfigurable architectures. FPGAs provide higher reconfigurability but with lower power and area efficiency due to the bit-level reconfiguration overhead. On the other hand, Coarse-Grained Reconfigurable Arrays (CGRAs) enables word-level reconfigurability with less reconfiguration overhead leading to better efficiency than FPGA~\cite{de2019coarse}. 

To achieve rapid dynamically reconfiguration, an ARENA node is prototyped with CGRA. The on-chip configuration memory is used for conserving the control signals for each task. The intra-node CGRA consists of 64 tiles connected in a mesh network. A scratchpad data memory conserves the data required for the computation. Note that both the control signals and the data are pre-loaded by the CPU/micro-controller through the DMA unit before launching the ARENA runtime. The CGRA communicates with the \texttt{Task Dispatcher} through the CGRA controller. The controller can offload tasks and coalesce spawned tokens through the \texttt{Coalescing Unit}.

\textbf{CGRA Tile --} As shown in Figure~\ref{fig:arena_node}, each CGRA tile contains a functional unit, a scratchpad control memory, a crossbar switch, and three sets of registers. The functional unit supports all the basic operations (e.g., $add$, $mul$, $shift$, $select$, $branch$, $load$, $store$, etc). Control-divergence (i.e., the existence of multiple control flow paths) inside the loop kernel is supported through partial predication~\cite{hamzeh2014branch}. The functional unit also supports the $spawn$ operation (i.e., generate a new task token and issue to the CGRA controller), which is unique in ARENA. If sufficient information is available ($TASK_{id}$, $TASK_{start}$ and $TASK_{end}$), a new token can be spawn in a single cycle; otherwise, two cycles are required to encode additional information (i.e., $PARAM$, $REMOTE_{start}$, and $REMOTE_{end}$). The $FROM_{node}$ filed will be automatically filled by the CGRA Controller. In Figure~\ref{fig:arena_node}, there are 4 tiles being able to spawn new tasks (marked in green). The leftmost tiles are connected to two 4-port scratchpad data memory banks. The functional unit can be configured to perform different operations at each cycle based on the control signals from the control memory. The CGRA tiles are granular to support the simultaneous execution of multiple tasks. Specifically, all the tiles are partitioned into 4 groups and a task can be executed by 1, 2, and 4 groups, dynamically managed by the CGRA controller.

\textbf{Control Memory --} The control signals of all the tasks are initially pre-loaded into the control memory. At runtime, tiles iterate over a subset of the control signals to execute specific tasks based on the $TASK_{id}$ in the task token. Each tile requires a 480-byte control memory in our prototype to support all application tasks evaluated in this paper (Section~\ref{sec:evaluation}). Each task has three execution modes powered by different tile groups. It takes only 8 cycles for the CGRA controller to reconfigure specific tile groups by using the data network to forward the $TASK_{id}$ systolically through the array from right to left.

%\todo[inline]{should mention sliding window?}

\begin{figure}[htb!]
	\centering
	\includegraphics[width=0.99\cw]{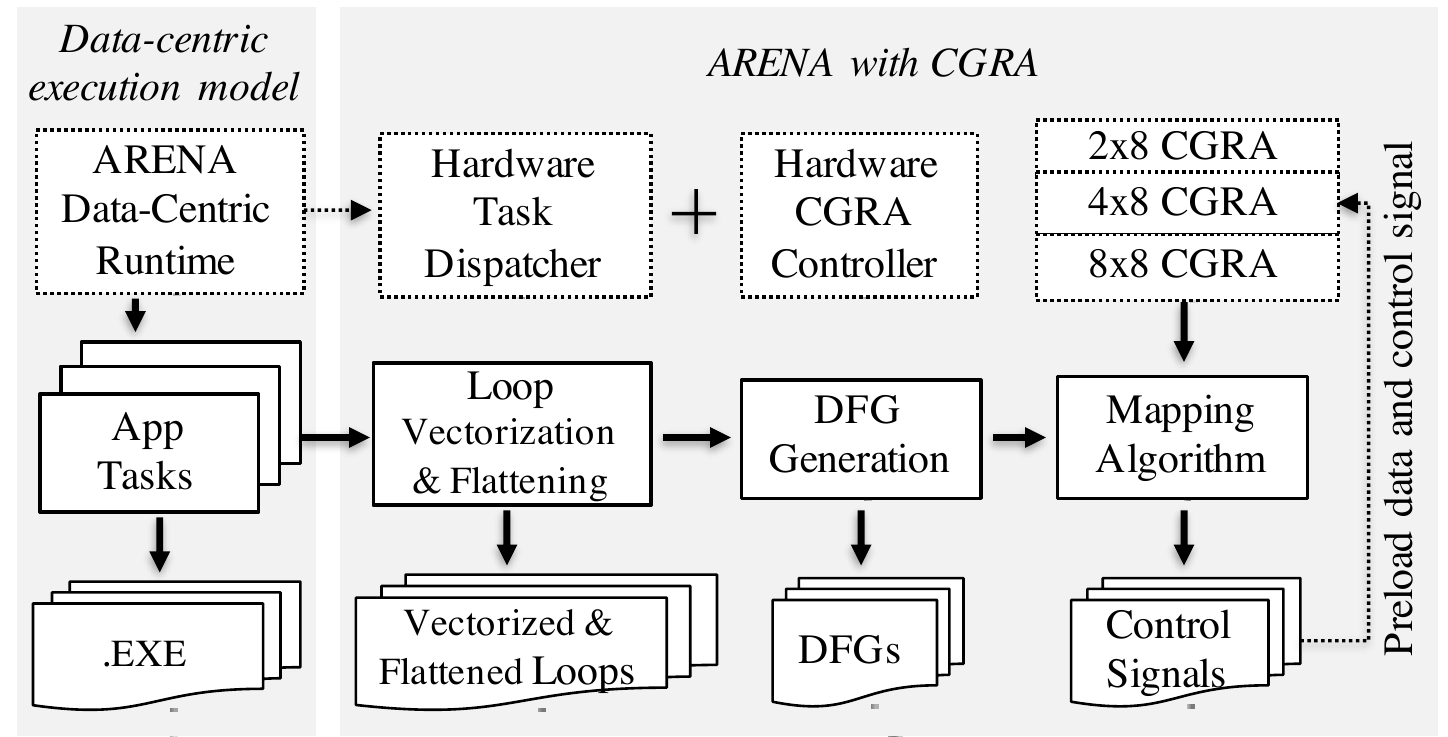}
	\caption{ARENA data-centric execution model development procedure with compiler toolchain.}
	\label{fig:procedure}
\end{figure}

\textbf{CGRA Controller --} The CGRA Controller can launch a task (using the task token at the head of \texttt{WaitQueue}) to be executed by different groups of the CGRA tiles. Based on the current CGRA utilization status and the data requirement of the target task, the CGRA controller allocate an appropriate number of groups for a waiting task. For example, if the data range required by the target task is less than a quarter of the local data range (i.e., $\text{TASK}_{\text{end}}$-$\text{TASK}_{\text{start}}<(\text{LOCAL}_{\text{end}}$-$\text{LOCAL}_{\text{start}}$)/4), only one available group (i.e., 2x8 CGRA tiles) will be allocated to the task. When the target task works on more than half of the local data range (i.e., $\text{TASK}_{\text{end}}$-$\text{TASK}_{\text{start}}>(\text{LOCAL}_{\text{end}}$-$\text{LOCAL}_{\text{start}}$)/2), the CGRA controller attempts to allocate all the four groups (i.e., entire 8x8 CGRA tiles) when available (otherwise, two groups are allocated). In addition, there are four queues in the controller to temporarily hold the spawned task tokens, which would be coalesced by the \texttt{Coalescing Unit} if any two of them imply continuous task data range and share the same $TASK_{id}$ and $PARAM$. When there are insufficient slots in the queues, the CGRA controller stops fetching tokens from the \texttt{WaitQueue} in the Task Dispatcher. Deadlock can be avoided by providing a memory attached to the CGRA controller for storing the over-spawned task tokens.

%An arbiter arbitrates one token out of four and issues it into the filter logic. 

%We use the \textit{TERMINATE} task to indicate the completion of an application to save the leakage power of the CGRA. To be specific, the power consumption of the reconfigurable accelerator in ARENA is composed of dynamic power and static power (i.e., leakage power). When a CGRA is executing a task, it is running in active mode with both dynamic and static power consumption. When a CGRA is in idle mode and waiting for a task, it consumes static power (i.e., leakage power)~\cite{hu2004microarchitectural}. Fortunately, the CGRA can even be switched off using power gating to save the leakage power When there is no task delivered in the ARENA ring. To indicate the completion of an application, \textit{TERMINATE} task is used. Whenever the CGRA completes the computation of a kernel and there is no task tag inside the send queue, a \textit{TERMINATE} task will be generated and pushed into the send queue. Both the receive queue and send queue discard the \textit{TERMINATE} task tag if the queue is not empty. When a task dispatcher encounters a \textit{TERMINATE} task tag for successive two times, the CGRA can be switched off.

\textbf{Compiler Toolchain for CGRA --} Figure~\ref{fig:procedure} shows the development procedure for ARENA using CGRA cluster as the backend. As already mentioned, the CGRA cluster is just one design choice; the ARENA execution model can be realized on alternative back-ends in case the HAF APIs (Table~\ref{tab:program_interface}) are realized. For example, on a CPU cluster backend, we can adopt MPI non-blocking-send primitives~\cite{gabriel2004open} to realize HAF APIs. We use this as one of our baselines in the evaluation. Here, to support the CGRA-cluster backend, an LLVM \cite{lattner2004llvm} based design automation toolchain is developed. In particular, a kernel that typically includes a multi-level nested loop is described as a task. As shown in Figure~\ref{fig:procedure}, to synthesize an appropriate mapping for a task on the allocated CGRA tiles, the nested loop is first vectorized with a factor that can fully leverage relatively larger CGRA tiles (e.g., 8x8 tiles) in the vectorization pass. Then, the remaining loops are flattened, generating the Control-Data Flow Graph (CDFG) representation, which is an extension of DFG with control dependence edges. We implemented a heuristic method~\cite{karunaratne2017hycube} to map the CDFG on various combinations of the tiles (i.e., 2x8, 4x8, and 8x8 tiles) and produce their control signals.

\section{Evaluation}\label{sec:evaluation}
\sloppy
%This section presents a detailed experimental evaluation of ARENA system architecture for its suitability in the next generation of data centers. We use high-level architectural simulations and RTL synthesis with representative workloads running in data centers for this evaluation.

%We evaluate ARENA in this section.

\subsection{Environment Setup}

%In this section, we present the experimental environment setup about the platforms and applications.

%\textbf{Platforms --} 
The ARENA runtime is evaluated on traditional CPU HPC clusters and our proposed CGRA-based ARENA cluster. For the latter, we extend the \emph{Structural Simulation Toolkit} (SST)~\cite{rodrigues2016structural} to model a multi-node cluster based on MPI. We model the network topology and package transmit switch in SST using the MACRELS Analytic Model~\cite{sstnetwork}. The token transmit network is modeled as 1D Torus Ring. We implement the task dispatcher, CGRA, and CGRA controller in PyMTL~\cite{lockhart2014pymtl}, which can report cycle-accurate simulation results for a single node. To obtain the cycle-level simulation result, we implement the dispatcher interface in SST to handle the task tokens. Finally, we feed the single-node result to SST and generates synthesizable Verilog for power, area, and timing analysis. The detailed simulation parameters are listed in Table \ref{tab:simulation_parameter}.

%with and without architectural support, respectively. Without architectural support, we evaluate the performance of compute-centric and data-centric execution of the applications running on CPU-only multi-node cluster. On the other hand, We evaluate the performance of ARNEA with architectural support by using a set of simulation tools.

%\rotatebox[origin=c]{90}{communication}

% \usepackage{multirow}
\begin{table}[htb!]
\centering
\footnotesize
\begin{tabular}{cl}
\hline\hline
  \textbf{Technology} &
  45nm \\ \hline
  \textbf{\begin{tabular}[c]{@{}c@{}}Network Interface\end{tabular}} &
  80 Gb/s \\ \hline
  \textbf{\begin{tabular}[c]{@{}c@{}}Network Topology\end{tabular}} &
  1D Torus Ring \\ \hline
   \textbf{\begin{tabular}[c]{@{}c@{}}Network Switch\end{tabular}} &
  \begin{tabular}[c]{@{}l@{}}1 per node, 1us hop latency\end{tabular} \\ \hline
  \textbf{Dispatcher} &
  \begin{tabular}[c]{@{}l@{}}Filter logic, 8-entry receive queue,
  8-entry \\wait queue, 8-entry send queue\end{tabular} \\ \hline
  \begin{tabular}[c]{@{}c@{}}\textbf{CPU} (baseline)\end{tabular} & \begin{tabular}[c]{@{}l@{}}2.6GHz, 20MB 3-level Cache, Out-of-order, x86\end{tabular} \\ \hline
  \textbf{CGRA} &
  \begin{tabular}[c]{@{}l@{}}8 $\times$ 8 CGRA, 480-byte control memory per tile, \\2-bank 4-port 32KB scratchpad data memory,\end{tabular} \\ \hline
  \textbf{\begin{tabular}[c]{@{}l@{}}CGRA Controller\end{tabular}} &
  \begin{tabular}[c]{@{}l@{}}4 $\times$ 4-entry queue for spawned task tokens, \\Coalescing unit\end{tabular} \\
  \hline\hline
\end{tabular}
\caption{RTL and simulation parameters for ARENA.}
\label{tab:simulation_parameter}
\end{table}

\textbf{Applications --} We evaluate ARENA using representative HPC and data-analytics workloads: The single-source shortest paths (\textbf{SSSP}) problem (see Section~\ref{subsec:programming}) is a key subroutine in many data-intensive graph computations.  General Matrix Multiply (\textbf{GEMM}) is the core function of linear algebra and deep learning workloads. We assume the matrices are distributed among nodes. Sparse-matrix-vector multiplication (\textbf{SPMV}) is the fundamental kernel in many scientific \& data applications. Here, the distributed matrix is in the Compressed Sparse Row (CSR) format. DNA sequence alignment (\textbf{DNA}) \cite{NBCIBLAST} leverages Needleman-Wunsch (NW) algorithm to search the best-matched protein sequences with respect to the target pattern. A Graph convolutional network \textbf{GCN}~\cite{fey2019fast} inference application on the Cora dataset is also evaluated, representing emerging irregular machine learning workloads. We assume the adjacency and feature matrices are distributed among nodes. Finally, an \textbf{N-body} simulation application \cite{green2010particle} for simulating dynamical particle system is demonstrated, representing traditional scientific simulation workloads. Again, the particle information (e.g., accelerations, velocity, position, collision, etc) are distributively stored and required to be updated per iteration at runtime.

%Running DNA on the distributed system with accelerator equipped, performance improvement can be achieved through pipelining but with limited speedup from loop-level parallelism constrained by the loop-carried dependency. 

%lower–upper (\textbf{LU}) decomposition kernel for solving square systems of linear equations is included in our evaluation. Note that the computation in LU kernel is performed in along the diagonal rather than row by row, which provides sufficient instruction-level parallelism for CGRA acceleration. 

The conventional compute-centric parallel implementations of all the evaluated applications are developed based on state-of-the-art algorithms or derived from the widely-used benchmark suites~\cite{pouchet2012polybench, che2009rodinia}. For example, the SSSP application is implemented based on~\cite{bulucc2011parallel}. The GCN model is extracted from PyTorch Geometric~\cite{fey2019fast}. The DNA application leverages the NW algorithm from Rodinia~\cite{che2009rodinia}. Regarding the ARENA implementation, all the applications are programmed following the ARENA programming models for data-centric asynchronous execution with runtime hardware specialization. Note that GCN and NBody contain numeric distinct functional tasks.

%Most applications and kernels feature with sufficient instruction-level and loop-level parallelism that are suitable to be accelerated by CGRA. 

\begin{figure*}[htb!]
	\centering
	\includegraphics[width=0.95\tw]{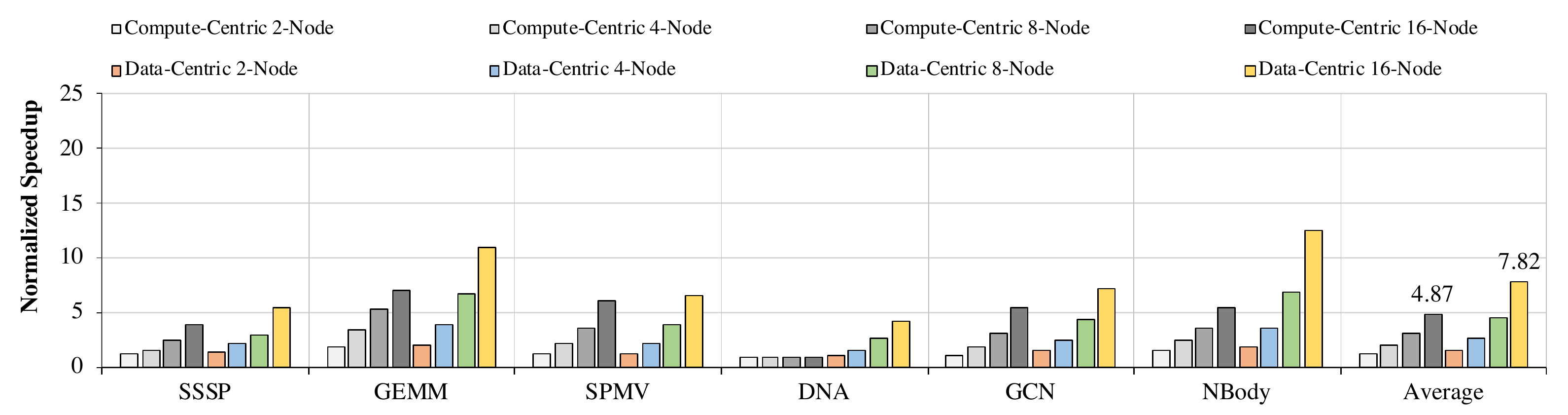}
	\caption{Normalized speedup for compute-centric and ARENA's data-centric execution models running on different multi-CPUs cluster w.r.t. a serial implementation on a single node.}
	\label{fig:data_centric_speedup}
\end{figure*}

\begin{figure}[htb!]
	\centering
	\includegraphics[width=0.95\cw]{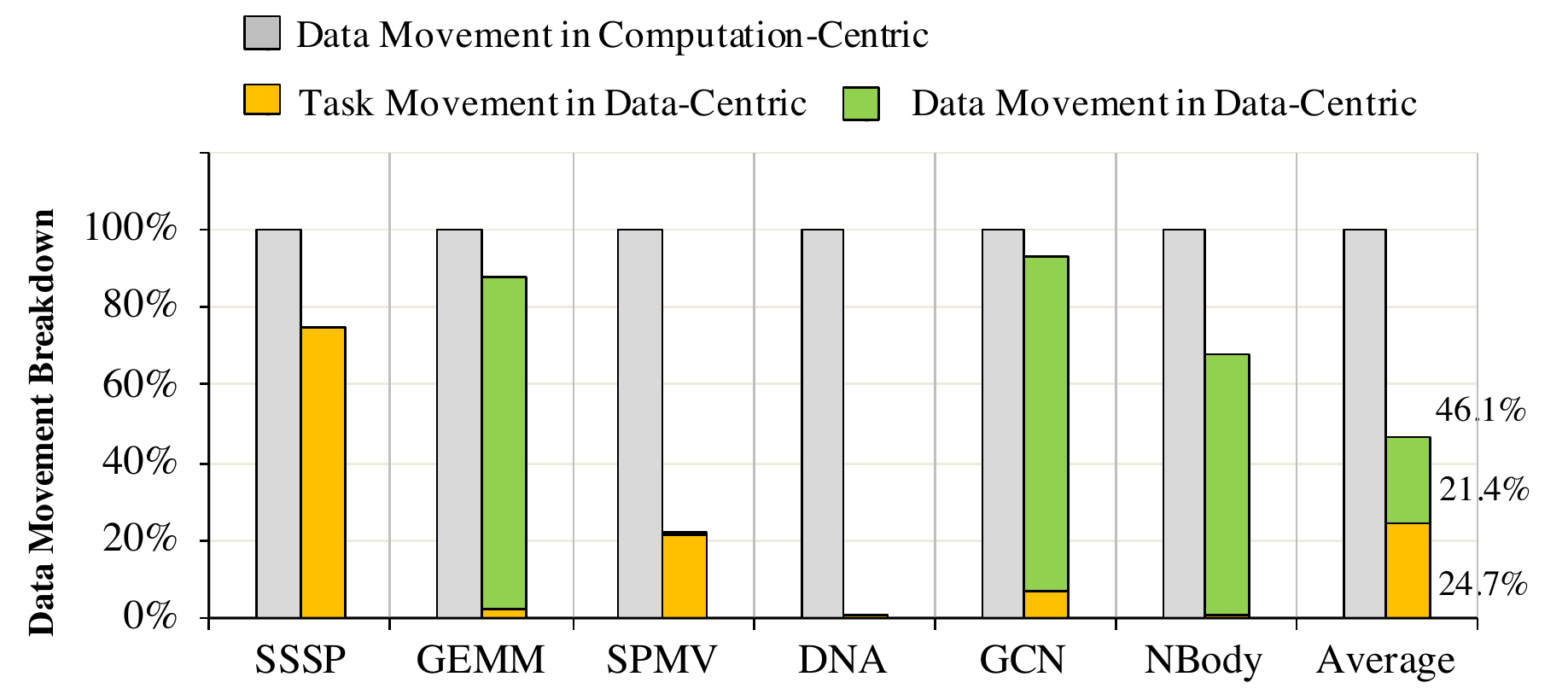}
	\caption{{Normalized data movement breakdown in data-centric model w.r.t. the compute-centric model.}}
	\label{fig:arena_utilization}
\end{figure}

\subsection{Evaluation Results}\label{subsec:results}

%Before diving into the performance improvement of ARENA system, we first show the performance improvements of ARENA's data-centric execution model and CGRA acceleration separately.
We show the benefits of the data-centric programming model, CGRA hardware acceleration, and the entire ARENA system.

\textbf{Programming Model --} We first show the performance effectiveness of ARENA's data-centric programming model. Figure~\ref{fig:data_centric_speedup} illustrates the normalized speedups for conventional compute-centric and ARENA's models (Both are software implemented based on MPI) with respect to a serial implementation on a single CPU node (i.e., baseline). As can be seen, ARENA's data-centric execution model shows higher speedups and better scalability in general. This is mainly due to the elimination of synchronization, and the minimization of data communication. On average, ARENA's data-centric model outperforms the compute-centric counterpart by 1.61$\times$ (i.e., 7.82/4.87) in a 16-node cluster. Specifically, the kernels with higher data parallelism (e.g. SSSP, GEMM, and SPMV) can gain better scalability for both models; for kernels with limited data parallelism such as DNA, the compute-centric model exhibits lower scalability due to massive data dependency and costly remote communication. In this condition, ARENA achieves better scalability by streaming task tokens over the nodes to minimize data movement.

%Even though there is limited data parallelism in the DNA application, the data-centric model achieves higher scalability by eliminating non-essential data communication. Specifically, compute-centric model of DNA is implemented based on an OpenMP model which all threads share the same global memory~\cite{che2009rodinia}. Each thread is assigned a sub-block of data from the global memory in a zig-zag manner and fetches/updates/writes back the required data block frequently, which incurs significant data movement. In ARENA, we observe that the data dependency only exits on the edge of the data block, which can be explicitly denoted by the parent task using the ARENA user APIs (i.e., $REMOTE_{start}$ and $REMOTE_{end}$ in ARENA\_task\_spawn()) and therefore minimizes the data movement.

%. Each iteration, active threads will fetch the required data block from the shared memory. Due to the data dependency, the updated data needs to be written back to the global memory at the end of the iteration. In this case, the frequent global memory copy incurs significant data movement if the local data block is stored on the remote node. In ARENA, we observe that the data dependency only exits on the edge of the data block, which significantly mitigates data movement.

\begin{figure*}[htb!]
	\centering
	\includegraphics[width=0.95\tw]{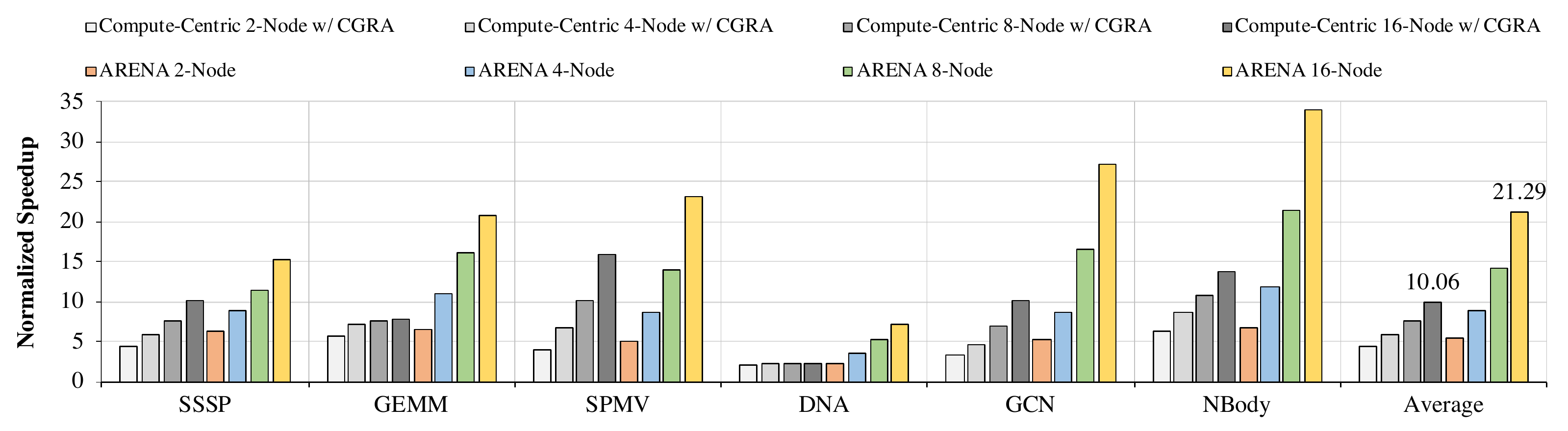}
	\caption{Normalized speedup for compute-centric and ARENA's data-centric execution models running on different multi-CGRAs cluster w.r.t. a serial implementation on a single node without any acceleration.}
	\label{fig:arena_speedup}
\end{figure*}

Figure~\ref{fig:arena_utilization} illustrates the normalized data movement breakdown for ARENA's data-centric model with respect to the compute-centric model for all applications in a 4-node cluster. Compared with the compute-centric HPC cluster, ARENA can eliminate on average 53.9\% data movement without any prior knowledge about the data distribution, leading to substantial improvement in energy efficiency. We also observe different data movement patterns across applications. For example, SSSP posts considerable task movement, as it spawns massive fine-grained tasks with discrete data-ranges, which are hard to coalesce. Regarding DNA, the compute-centric implementation is based on OpenMP where all threads are sharing the same copy in global memory~\cite{che2009rodinia}. The sub-blocks workload distribution to threads following a zig-zag manner incurs frequent data movement. In ARENA, the data dependency only exits on the edge of the sub-block, which can be explicitly labeled by the parent tasks using the ARENA User-APIs (i.e., $REMOTE_{start}$ and $REMOTE_{end}$ in ARENA\_task\_spawn()), therefore minimizing data movement.
Finally, GEMM and NBody comprise coarse-grained tasks and the task-flows require data streaming among the nodes, leading to little task movement or essential data movement as shown in the figure.
%We believe aggregating multiple tiny tasks into a larger one can significantly improve the performance and reduce data movement further, which will be explored in our future work. 
%The non-essential data movement of all the kernels is minimized. 

  %Specifically, each spawned task only works on a single row to detect its successors (see Figure~\ref{fig:bfs_arena_impl}).

\begin{figure}[htb!]
	\centering
	\includegraphics[width=0.95\cw]{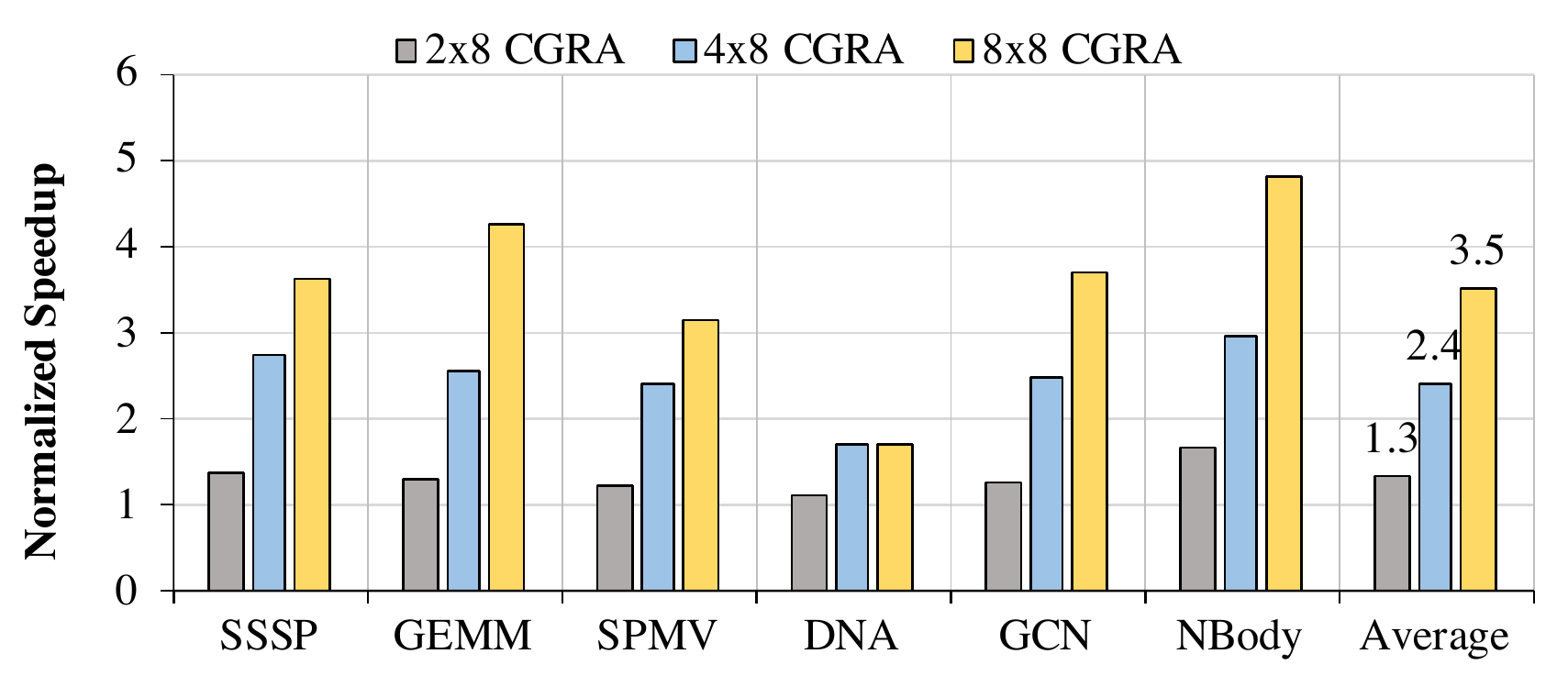}
	\caption{Normalized CGRA speedup w.r.t. the single node baseline CPU execution without any acceleration.}
	\label{fig:cgra_speedup}
\end{figure}

\textbf{CGRA Speedup --} Figure~\ref{fig:cgra_speedup} shows the normalized speedup of the evaluated applications running on different configurations or combinations of ARENA's CGRA tile groups (each group is a 2x8 CGRA) with respect to the single node CPU baseline. In general, a larger CGRA tile configuration leads to higher speedups. For DNA, however, the loop-carried data dependency limits the data parallelism, as well as the obtainable speedups (1.7$\times$ speedup at most). On average, 1.3$\times$, 2.4$\times$, and 3.5$\times$ speedups are achieved by ARENA's 2x8, 4x8, and 8x8 CGRA across all the applications and kernels. The 2x8 CGRA exhibits the optimal area-efficiency, showcasing the advantages of runtime hardware specialization.

\textbf{Overall System --} The normalized speedup of ARENA is shown in Figure~\ref{fig:arena_speedup}. As can be seen, ARENA maximizes the overall performance by leveraging a data-centric execution model and CGRA in a synergistic and integrated fashion. Instead of fixing the CGRA configuration for each workload, ARENA dynamically allocates and configures the CGRA tiles specifically for a particular task based on the task-carried specification (obtained based on data requirement of the task), as well as the current CGRA resource availability. On average, the compute-centric execution model using the entire CGRAs for each kernel obtains 10.06$\times$ speedup on a 16-node cluster, whereas  ARENA achieves 21.29$\times$ speedup. In other words, ARENA is 2.17$\times$ better than the compute-centric with CGRA support). This implies ARENA can leverage the CGRAs in a more efficient way. Compared with Figure~\ref{fig:data_centric_speedup}, we can see that ARENA with CGRAs also gains better scalability (from 1.61$\times$ to 2.17$\times$ on a 16-node cluster). The performance of DNA does not improve much due to limited acceleration from CGRA. Finally, the compute-centric execution of GEMM does not scale well because synchronization over a larger amount of data creates serious performance bottlenecks.

%On the other hand, asynchronous and streaming data movement still provides opportunities for ARENA to maintain the scalability on more than 16-node HPC cluster size.
%\todo[inline]{spmv though sparse, the final output is reduced. For MPI, it is a reduce on the entire V, for ARENA, it is sparse...}

\subsection{RTL Timing, Area and Power}

%To analyze the timing, area, and power statistics of ARENA's CGRA-cluster. 

\begin{figure}[htb!]
	\centering
	\includegraphics[width=0.99\cw]{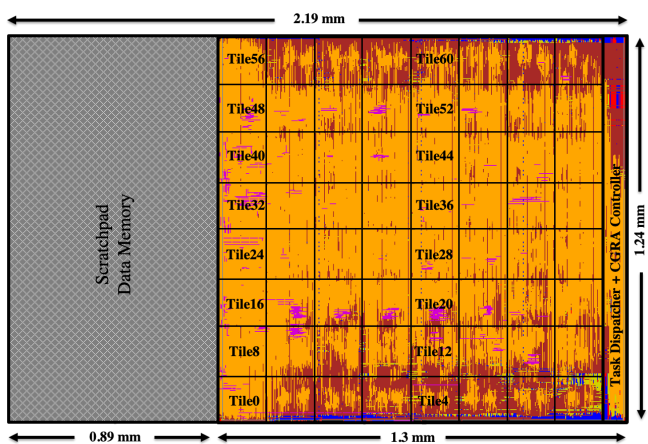}
	\caption{{Chip layout of a single ARENA node.}}
	\label{fig:layout}
\end{figure}

We evaluate the timing, area, and power consumption of ARENA's CGRA-cluster (e.g., CGRAs, CGRA controllers and task dispatchers) using the synthesized Verilog HDL code from PyMTL. We use Synopsys Design Compiler, Cadence Innovus, and Synopsys PrimeTime PX in order to synthesize, place, route, and estimate the power consumption of the designs. We use FreePDK45 with the Nangate standard cell library. Figure~\ref{fig:layout} illustrates the obtained chip layout. The area and power of the 32KB scratchpad data memory are estimated based on CACTI-6.5. The chip area is 2.19mm x 1.24mm and the operating frequency is 800MHz @ 45nm with on average 759.8mW power consumption.
%122 x 2 + 515.8 = 759.8

\section{Related Work}\label{sec:related}

We summarize related work regarding the ring network, clusters of reconfigurable architecture, and the task-based execution model in a ring network. 

%\todo[inline]{should we include all the task-based papers about different specific applications?}

\textbf{Ring Network.}
The ring network has been adopted in real multi-processors design \cite{barroso1993performance} (e.g., Intel Xeon-Phi \cite{chrysos2012intel}) and logically studied for efficient collective communication \cite{liu2015imr,awan2016efficient,alazemi2018routerless, li2019evaluating,wang2019achieving,lin2020deep}. On the one hand, the ring network provide simple routing mechanism to better-utilized the link bandwidth for fast communication \cite{barroso1993performance}. on the other hand, the ring network has been criticized for easy saturation due to linear increased latency with more node \cite{ainsworth2007characterizing, jerger2009chip, liu2015imr}. Previous works \cite{liu2015imr,alazemi2018routerless, wang2019achieving,lin2020deep} extend the ring concept to form local-rings network among a subset of nodes, known as routerless network. In ARENA, we avoid the saturation problem through: (a) a routerless task execution model among nodes; and (b) the dynamically task allocation and dispatching mechanism.

%Nodes in the network are typically connected by a unidirectional ring. Direct connection thus is only available among neighboring nodes. To reach a remote node, packets have to be injected and forwarded along the ring to the destination \cite{lin2020deep}. The single-ring network has a simple routing mechanism -- each node only needs to decide whether to take out a message from the ring or pass it along. Therefore, communication delays are shorter and the raw bandwidth provided by the link can be better-utilized \cite{barroso1993performance}. The ring concept has been extended for statically \cite{liu2015imr} or dynamically \cite{wang2019achieving} formed local-rings among a subset of nodes, known as routerless network \cite{alazemi2018routerless}. Although the single-ring network has been criticized for easy saturation due to linear increased latency with more nodes  \cite{ainsworth2007characterizing, jerger2009chip, liu2015imr}, this is not an issue for ARENA as: (a) No remote connection or intermediate routing is required based on our execution model; (b) We do not assume prior knowledge about data distribution, thus tasks have to be circulated around the ring anyway to locate their data.

\textbf{Reconfigurable Hardware Cluster.}
Clusters incorporating reconfigurable devices such as FPGAs \cite{putnam2014reconfigurable, dondo2015integrating, knodel2016rc3e, tarafdar2017enabling} and CGRAs \cite{prabhakar2017plasticine, jin2012implementation, HammerBlade} have already been showcased by existing works \cite{lyke2015introduction, tessier2015reconfigurable}. On one hand, Putnam et al. from Microsoft propose the reconfigurable Catapult fabric \cite{putnam2014reconfigurable}, where each instantiation consists of a 6x8 2D torus of Xilinx Stratix-V FPGAs. Every FPGA is connected to a CPU server via PCI-e and directly links to other FPGAs through the SAS cables. This 1632-node FPGA cluster has been adopted for document ranking of the Bing search engine. Nearly 95\% performance increase has been demonstrated with only a 10\% extra power budget. However, each FPGA in Catapult is specialized to a single application kernel during runtime. Reconfiguration takes several milliseconds. Even reloading models without changing the computation logic can take up to 250 microseconds. On the other hand, the data-centric execution model allowing different tasks working on the same set of data in an HPC node requires rapid dynamic reconfiguration. Gazzano et al. propose R-Grid, a complete grid infrastructure for distributed high-performance computing using dynamically reconfigurable FPGAs \cite{dondo2015integrating}. Knodel et al. virtualize the FPGA resources and propose adapted service models in a cloud context \cite{knodel2016rc3e}. Tarafdar et al. discussed how FPGAs of a cloud data center can be flexibly connected based on a logical kernel and a mapping file \cite{tarafdar2017enabling}. Zhang et al. adopt a cluster of six Xilinx VC709 FPGA for cooperative convolutional neural network inference \cite{zhang2016energy}.
Regarding to the CGRAs, the SambaNova DataScale system incorporates 8 reconfigurable-dataflow-units (RDUs) derived from Plasticine \cite{prabhakar2017plasticine}, claiming higher performance than a thousand GPUs for training extremely large deep-learning models. PPA~\cite{park2009polymorphic} exploits pipeline parallelism in streaming applications to create a CGRA-like pipeline to execute streaming media applications. Samsung proposes to adopt the CGRA cluster for medical volume image rendering \cite{jin2012implementation}. HammerBlade \cite{HammerBlade} aims at designing a rack-scale cluster for ML and Graphs. Their ASIC is composed of general-purpose cores and specialized CGRAs (e.g., Chimera \cite{ye2000chimaera}).

\textbf{Asynchronous Task Execution.}
The asynchronous tasks execution has been studied in many graph processing frameworks \cite{nasre2013data,nguyen2013lightweight,ben2017groute}.
They propose software-level data-centric approaches to implement irregular graph kernels on multi-node clusters \cite{nguyen2013lightweight} and GPUs platform \cite{nasre2013data}. The work that is most relevant to ARENA is Groute \cite{ben2017groute}, which is an asynchronous runtime environment for processing irregular graphs. In Groute, the GPUs form a logical ring network with each GPU conserves a worklist. Tasks are encapsulated as messages passing along the ring. ARENA is motivated by Groute. However, Groute is a pure software implementation based on general-purpose GPUs, thus cannot benefit from hardware specialization. Additionally, for Groute, the routing policy, which specifies the action when input is received, as well as memory consistency and ownership, are all defined and maintained by the users, which can be complicated, tedious and error-prone (e.g., imagine if data dependency occurs at runtime, users have to manually locate them and fetch in corresponding tasks). In ARENA, these are designed and supported by hardware. Furthermore, global coordination and work counting in Groute are centralized and managed by the CPU whereas in ARENA, all of them are distributedly processed.

\section{Conclusion}\label{sec:conclusion}
\sloppy

In this paper, we propose an asynchronous-reconfigurable-accelerator-ring architecture for next generation data-driven high-performance computing. Through the co-design of architecture and programming model, ARENA is able to bring computation tasks in the form of CGRA-specialized hardware accelerators to the data, rather than the reverse as in contemporary compute-centric and dataflow architectures, significantly improving performance and reducing data-movement. 

\section*{Acknowledgement}
This work was mainly supported by the Compute-Flow-Architecture project under PNNL's DMC LDRD Initiative. It was also supported by the SO$(\text{DA})^2$ and FALLACY projects under DMC. The evaluation platforms were supported by the U.S. DOE Office of Science, Office of Advanced Scientific Computing Research, under award 66150: "CENATE - Center for Advanced Architecture Evaluation". The Pacific Northwest National Laboratory is operated by Battelle for the U.S. Department of Energy under contract DE-AC05-76RL01830.

%-------------------------------------------------------------------------
% Back Matter
%-------------------------------------------------------------------------
\bibliographystyle{cbxabbrv}
\bibliography{ref}

\begin{thebibliography}{10}

\bibitem{nvlink_power}
{NVlink power.} Accessed April 10, 2020. https://rb.gy/ew3vf2.

\bibitem{Paderborn}
{Paderborn University Will Offer Intel CPU-FPGA Cluster for Researchers}.
\newblock \url{https://rb.gy/2vc51j}, 2017.

\bibitem{Genome}
{DOE Joint Genome Institute}.
\newblock \url{http://jgi.doe.gov}, 2018.

\bibitem{gridlab}
{GridLAB-D}.
\newblock \url{http://www.gridlabd.org}, 2018.

\bibitem{ExaCT}
{Exascale Combustion Co-Design Center (ExaCT)}.
\newblock \url{https://crd.lbl.gov/projects/combustion-codesign}, 2020.

\bibitem{NWChem}
{NWChem: Open Source High-Performance Computational Chemistry}.
\newblock \url{http://www.nwchem-sw.org}, 2020.

\bibitem{ainsworth2007characterizing}
T.~W. Ainsworth and T.~M. Pinkston.
\newblock On characterizing performance of the cell broadband engine element
  interconnect bus.
\newblock In {\em First International Symposium on Networks-on-Chip (NOCS'07)},
  pages 18--29. IEEE, 2007.

\bibitem{alazemi2018routerless}
F.~Alazemi, A.~Azizimazreah, B.~Bose, and L.~Chen.
\newblock Routerless network-on-chip.
\newblock In {\em 2018 IEEE International Symposium on High Performance
  Computer Architecture (HPCA)}, pages 492--503. IEEE, 2018.

\bibitem{alvaro2009boom}
P.~Alvaro, T.~Condie, N.~Conway, K.~Elmeleegy, J.~M. Hellerstein, and R.~C.
  Sears.
\newblock BOOM: Data-centric programming in the datacenter.
\newblock {\em EECS Department, University of California, Berkeley, Tech. Rep.
  UCB/EECS-2009-113}, 2009.

\bibitem{awan2016efficient}
A.~A. Awan, K.~Hamidouche, A.~Venkatesh, and D.~K. Panda.
\newblock Efficient large message broadcast using NCCL and CUDA-aware MPI for
  deep learning.
\newblock In {\em Proceedings of the 23rd European MPI Users' Group Meeting},
  pages 15--22, 2016.

\bibitem{barroso1993performance}
L.~A. Barroso and M.~Dubois.
\newblock The performance of cache-coherent ring-based multiprocessors.
\newblock In {\em Proceedings of the 20th annual international symposium on
  computer architecture}, pages 268--277, 1993.

\bibitem{bauer2012legion}
M.~Bauer, S.~Treichler, E.~Slaughter, and A.~Aiken.
\newblock Legion: Expressing locality and independence with logical regions.
\newblock In {\em SC'12: Proceedings of the International Conference on High
  Performance Computing, Networking, Storage and Analysis}, pages 1--11. IEEE,
  2012.

\bibitem{ben2017groute}
T.~Ben-Nun, M.~Sutton, S.~Pai, and K.~Pingali.
\newblock {Groute: An Asynchronous Multi-GPU Programming Model for Irregular
  Computations}.
\newblock {\em ACM SIGPLAN Notices}, 52(8):235--248, 2017.

\bibitem{bisseling1993scientific}
R.~H. Bisseling and W.~F. McColl.
\newblock Scientific computing on bulk synchronous parallel architectures.
\newblock 1993.

\bibitem{blumofe1996cilk}
R.~D. Blumofe, C.~F. Joerg, B.~C. Kuszmaul, C.~E. Leiserson, K.~H. Randall, and
  Y.~Zhou.
\newblock Cilk: An efficient multithreaded runtime system.
\newblock {\em Journal of parallel and distributed computing}, 37(1):55--69,
  1996.

\bibitem{bonachea2017gasnet}
D.~Bonachea and P.~Hargrove.
\newblock GASNet Specification, v1. 8.1.
\newblock 2017.

\bibitem{bonachea2018gasnet}
D.~Bonachea and P.~H. Hargrove.
\newblock GASNet-EX: A high-performance, portable communication library for
  exascale.
\newblock In {\em International Workshop on Languages and Compilers for
  Parallel Computing}, pages 138--158. Springer, 2018.

\bibitem{boroumand2018google}
A.~Boroumand, S.~Ghose, Y.~Kim, R.~Ausavarungnirun, E.~Shiu, R.~Thakur, D.~Kim,
  A.~Kuusela, A.~Knies, P.~Ranganathan, and O.~Mutlu.
\newblock Google Workloads for Consumer Devices: Mitigating Data Movement
  Bottlenecks.
\newblock In {\em Proceedings of the Twenty-Third International Conference on
  Architectural Support for Programming Languages and Operating Systems}, pages
  316--331, 2018.

\bibitem{bulucc2011parallel}
A.~Bulu{\c{c}} and K.~Madduri.
\newblock Parallel breadth-first search on distributed memory systems.
\newblock In {\em Proceedings of 2011 International Conference for High
  Performance Computing, Networking, Storage and Analysis}, pages 1--12, 2011.

\bibitem{chamberlain2007parallel}
B.~L. Chamberlain, D.~Callahan, and H.~P. Zima.
\newblock Parallel programmability and the chapel language.
\newblock {\em The International Journal of High Performance Computing
  Applications}, 21(3):291--312, 2007.

\bibitem{charles2005x10}
P.~Charles, C.~Grothoff, V.~Saraswat, C.~Donawa, A.~Kielstra, K.~Ebcioglu,
  C.~Von~Praun, and V.~Sarkar.
\newblock X10: an object-oriented approach to non-uniform cluster computing.
\newblock {\em Acm Sigplan Notices}, 40(10):519--538, 2005.

\bibitem{che2009rodinia}
S.~Che, M.~Boyer, J.~Meng, D.~Tarjan, J.~W. Sheaffer, S.-H. Lee, and
  K.~Skadron.
\newblock Rodinia: A benchmark suite for heterogeneous computing.
\newblock In {\em 2009 IEEE international symposium on workload
  characterization (IISWC)}, pages 44--54. Ieee, 2009.

\bibitem{chrysos2012intel}
G.~Chrysos.
\newblock Intel{\textregistered} xeon phi coprocessor (codename knights
  corner).
\newblock In {\em 2012 IEEE Hot Chips 24 Symposium (HCS)}, pages 1--31. IEEE,
  2012.

\bibitem{de2019coarse}
B.~De~Sutter, P.~Raghavan, and A.~Lambrechts.
\newblock Coarse-grained reconfigurable array architectures.
\newblock In {\em Handbook of signal processing systems}, pages 427--472.
  Springer, 2019.

\bibitem{dean2008mapreduce}
J.~Dean and S.~Ghemawat.
\newblock MapReduce: simplified data processing on large clusters.
\newblock {\em Communications of the ACM}, 51(1):107--113, 2008.

\bibitem{dondo2015integrating}
J.~Dondo~Gazzano, F.~Sanchez~Molina, F.~Rincon, and J.~C. L{\'o}pez.
\newblock Integrating reconfigurable hardware-based grid for high performance
  computing.
\newblock {\em The Scientific World Journal}, 2015, 2015.

\bibitem{dongarra2020numerical}
J.~Dongarra, L.~Grigori, and N.~J. Higham.
\newblock Numerical algorithms for high-performance computational science.
\newblock {\em Philosophical Transactions of the Royal Society A},
  378(2166):20190066, 2020.

\bibitem{fey2019fast}
M.~Fey and J.~E. Lenssen.
\newblock Fast graph representation learning with PyTorch Geometric.
\newblock {\em arXiv preprint arXiv:1903.02428}, 2019.

\bibitem{gabriel2004open}
E.~Gabriel, G.~E. Fagg, G.~Bosilca, T.~Angskun, J.~J. Dongarra, J.~M. Squyres,
  V.~Sahay, P.~Kambadur, B.~Barrett, A.~Lumsdaine, R.~H. Castain, D.~J. Daniel,
  R.~L. Graham, and T.~S. Woodall.
\newblock Open MPI: Goals, concept, and design of a next generation MPI
  implementation.
\newblock In D.~Kranzlm{\"u}ller, P.~Kacsuk, and J.~Dongarra, editors, {\em
  European Parallel Virtual Machine/Message Passing Interface Users’ Group
  Meeting}, pages 97--104. Springer, 2004.

\bibitem{geng2018fpdeep}
T.~Geng, T.~Wang, A.~Sanaullah, C.~Yang, R.~Xu, R.~Patel, and M.~Herbordt.
\newblock FPDeep: Acceleration and load balancing of CNN training on FPGA
  clusters.
\newblock In {\em 2018 IEEE 26th Annual International Symposium on
  Field-Programmable Custom Computing Machines (FCCM)}, pages 81--84. IEEE,
  2018.

\bibitem{green2010particle}
S.~Green.
\newblock Particle simulation using cuda.
\newblock {\em NVIDIA whitepaper}, 6:121--128, 2010.

\bibitem{hamzeh2014branch}
M.~Hamzeh, A.~Shrivastava, and S.~Vrudhula.
\newblock Branch-aware loop mapping on CGRAs.
\newblock In {\em Proceedings of the 51st Annual Design Automation Conference},
  pages 1--6, 2014.

\bibitem{hazelwood2018applied}
K.~{Hazelwood}, S.~{Bird}, D.~{Brooks}, S.~{Chintala}, U.~{Diril},
  D.~{Dzhulgakov}, M.~{Fawzy}, B.~{Jia}, Y.~{Jia}, A.~{Kalro}, J.~{Law},
  K.~{Lee}, J.~{Lu}, P.~{Noordhuis}, M.~{Smelyanskiy}, L.~{Xiong}, and
  X.~{Wang}.
\newblock Applied machine learning at facebook: A datacenter infrastructure
  perspective.
\newblock In {\em 2018 IEEE International Symposium on High Performance
  Computer Architecture (HPCA)}, pages 620--629. IEEE, 2018.

\bibitem{horowitz20141}
M.~Horowitz.
\newblock 1.1 computing's energy problem (and what we can do about it).
\newblock In {\em 2014 IEEE International Solid-State Circuits Conference
  Digest of Technical Papers (ISSCC)}, pages 10--14. IEEE, 2014.

\bibitem{jerger2009chip}
N.~E. Jerger and L.-S. Peh.
\newblock On-chip networks.
\newblock {\em Synthesis Lectures on Computer Architecture}, 4(1):1--141, 2009.

\bibitem{jin2012implementation}
S.~Jin, S.~Lee, M.-K. Chung, Y.~Cho, and S.~Ryu.
\newblock Implementation of a volume rendering on coarse-grained reconfigurable
  multiprocessor.
\newblock In {\em 2012 International Conference on Field-Programmable
  Technology}, pages 243--246. IEEE, 2012.

\bibitem{NBCIBLAST}
M.~Johnson, I.~Zaretskaya, Y.~Raytselis, Y.~Merezhuk, S.~McGinnis, and T.~L.
  Madden.
\newblock {NCBI BLAST: a better web interface}.
\newblock volume~36, pages W5--W9, 04 2008.

\bibitem{jouppi2017datacenter}
N.~P. {Jouppi}, C.~{Young}, N.~{Patil}, D.~{Patterson}, G.~{Agrawal},
  R.~{Bajwa}, S.~{Bates}, S.~{Bhatia}, N.~{Boden}, A.~{Borchers}, R.~{Boyle},
  P.~{Cantin}, C.~{Chao}, C.~{Clark}, J.~{Coriell}, M.~{Daley}, M.~{Dau},
  J.~{Dean}, B.~{Gelb}, T.~V. {Ghaemmaghami}, R.~{Gottipati}, W.~{Gulland},
  R.~{Hagmann}, C.~R. {Ho}, D.~{Hogberg}, J.~{Hu}, R.~{Hundt}, D.~{Hurt},
  J.~{Ibarz}, A.~{Jaffey}, A.~{Jaworski}, A.~{Kaplan}, H.~{Khaitan},
  D.~{Killebrew}, A.~{Koch}, N.~{Kumar}, S.~{Lacy}, J.~{Laudon}, J.~{Law},
  D.~{Le}, C.~{Leary}, Z.~{Liu}, K.~{Lucke}, A.~{Lundin}, G.~{MacKean},
  A.~{Maggiore}, M.~{Mahony}, K.~{Miller}, R.~{Nagarajan}, R.~{Narayanaswami},
  R.~{Ni}, K.~{Nix}, T.~{Norrie}, M.~{Omernick}, N.~{Penukonda}, A.~{Phelps},
  J.~{Ross}, M.~{Ross}, A.~{Salek}, E.~{Samadiani}, C.~{Severn}, G.~{Sizikov},
  M.~{Snelham}, J.~{Souter}, D.~{Steinberg}, A.~{Swing}, M.~{Tan},
  G.~{Thorson}, B.~{Tian}, H.~{Toma}, E.~{Tuttle}, V.~{Vasudevan}, R.~{Walter},
  W.~{Wang}, E.~{Wilcox}, and D.~H. {Yoon}.
\newblock In-datacenter performance analysis of a tensor processing unit.
\newblock In {\em 2017 ACM/IEEE 44th Annual International Symposium on Computer
  Architecture (ISCA)}, pages 1--12, 2017.

\bibitem{karunaratne2017hycube}
M.~Karunaratne, A.~K. Mohite, T.~Mitra, and L.-S. Peh.
\newblock Hycube: A cgra with reconfigurable single-cycle multi-hop
  interconnect.
\newblock In {\em Proceedings of the 54th Annual Design Automation Conference
  2017}, pages 1--6, 2017.

\bibitem{kestor2013quantifying}
G.~Kestor, R.~Gioiosa, D.~J. Kerbyson, and A.~Hoisie.
\newblock Quantifying the energy cost of data movement in scientific
  applications.
\newblock In {\em 2013 IEEE international symposium on workload
  characterization (IISWC)}, pages 56--65. IEEE, 2013.

\bibitem{knodel2016rc3e}
O.~Knodel, P.~Lehmann, and R.~G. Spallek.
\newblock RC3E: Reconfigurable accelerators in data centres and their provision
  by adapted service models.
\newblock In {\em 2016 IEEE 9th International Conference on Cloud Computing
  (CLOUD)}, pages 19--26. IEEE, 2016.

\bibitem{lattner2004llvm}
C.~Lattner and V.~Adve.
\newblock LLVM: A compilation framework for lifelong program analysis \&
  transformation.
\newblock In {\em International Symposium on Code Generation and Optimization,
  2004. CGO 2004.}, pages 75--86. IEEE, 2004.

\bibitem{lee2017introducing}
K.~Lee.
\newblock Introducing big basin: Our next-generation ai hardware, 2017.

\bibitem{li2019evaluating}
A.~Li, S.~L. Song, J.~Chen, J.~Li, X.~Liu, N.~R. Tallent, and K.~J. Barker.
\newblock Evaluating Modern GPU Interconnect: PCIe, NVLink, NV-SLI, NVSwitch
  and GPUDirect.
\newblock {\em IEEE Transactions on Parallel and Distributed Systems},
  31(1):94--110, 2019.

\bibitem{lin2020deep}
T.-R. Lin, D.~Penney, M.~Pedram, and L.~Chen.
\newblock A Deep Reinforcement Learning Framework for Architectural
  Exploration: A Routerless NoC Case Study.
\newblock In {\em 2020 IEEE International Symposium on High Performance
  Computer Architecture (HPCA)}. IEEE, 2018.

\bibitem{liu2015imr}
S.~Liu, T.~Chen, L.~Li, X.~Feng, Z.~Xu, H.~Chen, F.~Chong, and Y.~Chen.
\newblock IMR: High-performance low-cost multi-ring NoCs.
\newblock {\em IEEE Transactions on Parallel and Distributed Systems},
  27(6):1700--1712, 2015.

\bibitem{lockhart2014pymtl}
D.~Lockhart, G.~Zibrat, and C.~Batten.
\newblock PyMTL: A unified framework for vertically integrated computer
  architecture research.
\newblock In {\em 2014 47th Annual IEEE/ACM International Symposium on
  Microarchitecture}, pages 280--292. IEEE, 2014.

\bibitem{lyke2015introduction}
J.~C. Lyke, C.~G. Christodoulou, G.~A. Vera, and A.~H. Edwards.
\newblock An introduction to reconfigurable systems.
\newblock {\em Proceedings of the IEEE}, 103(3):291--317, 2015.

\bibitem{Seismic}
P.~J. Maechling.
\newblock {Exascale Applications in Seismic Hazard Analysis}.
\newblock
  \url{http://www.nics.tennessee.edu/files/pdf/Maechling_Exascale_072910_v9.pdf},
  2018.

\bibitem{molka2010characterizing}
D.~Molka, D.~Hackenberg, R.~Sch{\"o}ne, and M.~S. M{\"u}ller.
\newblock Characterizing the energy consumption of data transfers and
  arithmetic operations on x86- 64 processors.
\newblock In {\em International conference on green computing}, pages 123--133.
  IEEE, 2010.

\bibitem{moore1965cramming}
G.~E. Moore.
\newblock Cramming more components onto integrated circuits, 1965.

\bibitem{nasre2013data}
R.~Nasre, M.~Burtscher, and K.~Pingali.
\newblock Data-driven versus topology-driven irregular computations on GPUs.
\newblock In {\em 2013 IEEE 27th International Symposium on Parallel and
  Distributed Processing}, pages 463--474. IEEE, 2013.

\bibitem{nguyen2013lightweight}
D.~Nguyen, A.~Lenharth, and K.~Pingali.
\newblock A lightweight infrastructure for graph analytics.
\newblock In {\em Proceedings of the Twenty-Fourth ACM Symposium on Operating
  Systems Principles}, pages 456--471, 2013.

\bibitem{nvidia2017dgx1}
NVIDIA.
\newblock {NVIDIA DGX-1 System Architecture White Paper}, 2017.

\bibitem{ohmura2014mdgrape}
I.~Ohmura, G.~Morimoto, Y.~Ohno, A.~Hasegawa, and M.~Taiji.
\newblock MDGRAPE-4: a special-purpose computer system for molecular dynamics
  simulations.
\newblock {\em Philosophical Transactions of the Royal Society A: Mathematical,
  Physical and Engineering Sciences}, 372(2021):20130387, 2014.

\bibitem{GROMACS}
S.~P{\'a}ll, M.~J. Abraham, C.~Kutzner, B.~Hess, and E.~Lindahl.
\newblock Tackling Exascale Software Challenges in Molecular Dynamics
  Simulations with GROMACS.
\newblock In S.~Markidis and E.~Laure, editors, {\em Solving Software
  Challenges for Exascale}, pages 3--27, Cham, 2015. Springer International
  Publishing.

\bibitem{park2009polymorphic}
H.~Park, Y.~Park, and S.~Mahlke.
\newblock Polymorphic pipeline array: a flexible multicore accelerator with
  virtualized execution for mobile multimedia applications.
\newblock In {\em Proceedings of the 42nd Annual IEEE/ACM International
  Symposium on Microarchitecture}, pages 370--380, 2009.

\bibitem{perez2010handling}
J.~M. Perez, R.~M. Badia, and J.~Labarta.
\newblock Handling task dependencies under strided and aliased references.
\newblock In {\em Proceedings of the 24th ACM International Conference on
  Supercomputing}, pages 263--274, 2010.

\bibitem{pouchet2012polybench}
L.-N. Pouchet and S.~Grauer-Gray.
\newblock Polybench: The polyhedral benchmark suite.
\newblock {\em URL: http://www. cs. ucla. edu/pouchet/software/polybench},
  2012.

\bibitem{prabhakar2017plasticine}
R.~Prabhakar, Y.~Zhang, D.~Koeplinger, M.~Feldman, T.~Zhao, S.~Hadjis,
  A.~Pedram, C.~Kozyrakis, and K.~Olukotun.
\newblock Plasticine: A reconfigurable architecture for parallel patterns.
\newblock In {\em 2017 ACM/IEEE 44th Annual International Symposium on Computer
  Architecture (ISCA)}, pages 389--402. IEEE, 2017.

\bibitem{putnam2014reconfigurable}
A.~Putnam, A.~M. Caulfield, E.~S. Chung, D.~Chiou, K.~Constantinides, J.~Demme,
  H.~Esmaeilzadeh, J.~Fowers, G.~P. Gopal, J.~Gray, M.~Haselman, S.~Hauck,
  S.~Heil, A.~Hormati, J.-Y. Kim, S.~Lanka, J.~Larus, E.~Peterson, S.~Pope,
  A.~Smith, J.~Thong, P.~Y. Xiao, and D.~Burger.
\newblock A reconfigurable fabric for accelerating large-scale datacenter
  services.
\newblock In {\em 2014 ACM/IEEE 41st International Symposium on Computer
  Architecture (ISCA)}, pages 13--24. IEEE, 2014.

\bibitem{rodrigues2016structural}
A.~F. Rodrigues, G.~R. Voskuilen, S.~D. Hammond, and K.~S. Hemmert.
\newblock Structural Simulation Toolkit (SST).
\newblock Technical report, Sandia National Lab.(SNL-NM), Albuquerque, NM
  (United States), 2016.

\bibitem{shalf2020future}
J.~Shalf.
\newblock The future of computing beyond Moore’s law.
\newblock {\em Philosophical Transactions of the Royal Society A},
  378(2166):20190061, 2020.

\bibitem{shaw2014anton}
D.~E. Shaw, J.~Grossman, J.~A. Bank, B.~Batson, J.~A. Butts, J.~C. Chao, M.~M.
  Deneroff, R.~O. Dror, A.~Even, C.~H. Fenton, et~al.
\newblock Anton 2: raising the bar for performance and programmability in a
  special-purpose molecular dynamics supercomputer.
\newblock In {\em SC'14: Proceedings of the International Conference for High
  Performance Computing, Networking, Storage and Analysis}, pages 41--53. IEEE,
  2014.

\bibitem{tarafdar2017enabling}
N.~Tarafdar, T.~Lin, E.~Fukuda, H.~Bannazadeh, A.~Leon-Garcia, and P.~Chow.
\newblock Enabling flexible network FPGA clusters in a heterogeneous cloud data
  center.
\newblock In {\em Proceedings of the 2017 ACM/SIGDA International Symposium on
  Field-Programmable Gate Arrays}, pages 237--246, 2017.

\bibitem{HammerBlade}
M.~B. Taylor, A.~Sampson, C.~Batten, Z.~Zhang, L.~Ceze, M.~Oskin, and
  D.~Richmond.
\newblock The HammerBlade: An ML-Optimized Supercomputer for ML and Graphs.
\newblock
  \url{https://sampl.cs.washington.edu/tvmconf/slides/Michael-Taylor-HammerBlade.pdf},
  2020.

\bibitem{tessier2015reconfigurable}
R.~Tessier, K.~Pocek, and A.~DeHon.
\newblock Reconfigurable computing architectures.
\newblock {\em Proceedings of the IEEE}, 103(3):332--354, 2015.

\bibitem{Carbon}
D.~Trebotich, M.~F. Adams, S.~Molins, C.~I. Steefel, and C.~Shen.
\newblock High-Resolution Simulation of Pore-Scale Reactive Transport Processes
  Associated with Carbon Sequestration.
\newblock {\em Computing in Science Engineering}, 16(6):22--31, Nov 2014.

\bibitem{valiant1990bridging}
L.~G. Valiant.
\newblock A bridging model for parallel computation.
\newblock {\em Communications of the ACM}, 33(8):103--111, 1990.

\bibitem{sstnetwork}
F.~Versolatto and A.~Tonello.
\newblock An MTL Theory Approach for the Simulation of MIMO Power-Line
  Communication Channels.
\newblock {\em Power Delivery, IEEE Transactions on}, 26:1710--1717, 07 2011.

\bibitem{vetter2018extreme}
J.~S. Vetter, R.~Brightwell, M.~Gokhale, P.~McCormick, R.~Ross, J.~Shalf,
  K.~Antypas, D.~Donofrio, T.~Humble, C.~Schuman, et~al.
\newblock Extreme heterogeneity 2018-productive computational science in the
  era of extreme heterogeneity: Report for DOE ASCR workshop on extreme
  heterogeneity.
\newblock Technical report, USDOE Office of Science (SC), Washington, DC
  (United States), 2018.

\bibitem{wang2019achieving}
L.~Wang, L.~Liu, J.~Han, X.~Wang, S.~Yin, and S.~Wei.
\newblock Achieving Flexible Global Reconfiguration in NoCs using
  Reconfigurable Rings.
\newblock {\em IEEE Transactions on Parallel and Distributed Systems}, 2019.

\bibitem{quantummech}
L.-W. Wang.
\newblock Divide-and-conquer quantum mechanical material simulations with
  exascale supercomputers.
\newblock {\em National Science Review}, 1(4):604--617, 2014.

\bibitem{ye2000chimaera}
Z.~A. Ye, A.~Moshovos, S.~Hauck, and P.~Banerjee.
\newblock CHIMAERA: a high-performance architecture with a tightly-coupled
  reconfigurable functional unit.
\newblock {\em ACM SIGARCH Computer Architecture News}, 28(2):225--235, 2000.

\bibitem{zhang2016energy}
C.~Zhang, D.~Wu, J.~Sun, G.~Sun, G.~Luo, and J.~Cong.
\newblock Energy-efficient CNN implementation on a deeply pipelined FPGA
  cluster.
\newblock In {\em Proceedings of the 2016 International Symposium on Low Power
  Electronics and Design}, pages 326--331, 2016.

\end{thebibliography}
\end{document}